\documentclass[10pt,aps,prd,reprint,nofootinbib,floats,floatfix,amsfonts,amssymb,amsmath,preprintnumbers,notitlepage,superscriptaddress,twocolumn]{revtex4-1}

\usepackage{graphicx,color} 
\usepackage{hyphenat} 
\usepackage{amsmath, amssymb, amsfonts,amsbsy,amsthm} 
\usepackage{float}
\usepackage{textcomp}
\usepackage{subcaption}
\usepackage{soul,xcolor} 
\setstcolor{red}

\newcommand{\orcid}[1]{\href{https://orcid.org/#1}{\includegraphics[scale=0.15]{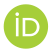}}}
\newcommand{\Kappa}[0]{\scalebox{1.5}{$\kappa$}}

\usepackage[normalem]{ulem}

\begin{document}

\title{Impact of correlated seismic and correlated Newtonian noise on the Einstein Telescope}

\author{Kamiel Janssens\orcid{0000-0001-8760-4429}}
\affiliation{Universiteit Antwerpen, Prinsstraat 13, 2000 Antwerpen, Belgium}
\affiliation{Artemis, Universit\'{e} C\^{o}te d’Azur, Observatoire C\^{o}te d’Azur, Nice, France}
\author{Guillaume Boileau \orcid{0000-0002-3576-69689}}
\affiliation{Universiteit Antwerpen, Prinsstraat 13, 2000 Antwerpen, Belgium}
\author{Nelson Christensen\orcid{0000-0002-6870-4202}}
\affiliation{Artemis, Universit\'{e} C\^{o}te d’Azur, Observatoire C\^{o}te d’Azur, Nice, France}
\author{Francesca Badaracco \orcid{0000-0001-8553-7904}}
\affiliation{Universit\'{e} Catholique de Louvain, Louvain-La-Neuve, Belgium}
\author{Nick van Remortel \orcid{0000-0003-4180-8199}}
\affiliation{Universiteit Antwerpen, Prinsstraat 13, 2000 Antwerpen, Belgium}

\date{\today}

\begin{abstract}

Correlated noise could impact the search for the gravitational wave background at future Earth-based gravitational-wave detectors. Due to the small distance ($\sim$ 400\,m) between the different interferometers of the Einstein Telescope, correlated seismic noise could have a significant effect.
To this extent, we study the seismic correlations at the Earth's surface, as well as underground, between seismometers and geophones separated by several hundreds of meters, in the frequency range 0.05\,Hz - 50\,Hz.
Based on these correlated seismic fields we predict the levels of correlated Newtonian noise (NN). 
We construct upper limits on the allowed seismic coupling function such that correlated seismic noise does not affect the search for an isotropic gravitational wave background.
Assuming a facility located 300\,m below the surface, the impact on the search for a gravitational wave background of correlated NN from Rayleigh waves are found to be problematic up to $\sim$ 5\,Hz. The NN from body waves, however, constitutes a serious threat to the search of a gravitational wave background. Correlated NN from body waves could be up to five to seven orders of magnitude above the planned sensitivity at $\sim$ 3\,Hz and it could impede any search for a gravitational wave background below 40\,Hz. With a factor 10 of NN reduction via NN cancellation in each interferometer, the effects of the NN on the stochastic search could be eliminated above 30\,Hz.  

\end{abstract}

\maketitle


\section{Introduction}
\label{sec:Introduction}

Searches for an isotropic gravitational-wave background (GWB)~\cite{Christensen_2018} typically rely on cross-correlating data from two (or more) interferometers. In the case of the current ($\rm 2^{\rm nd}$) generation of Earth-based gravitational wave (GW) interferometers -- LIGO \cite{2015}, Virgo \cite{VIRGO:2014yos} and KAGRA \cite{PhysRevD.88.043007} -- they are separated by thousands of kilometers. This large separation is a very effective way to reduce the amount of correlated noise between the different interferometers. One source which is known to be correlated over such long distances are the Schumann resonances~\cite{Schumann1,Schumann2}. They have been extensively studied in the context of LIGO, Virgo and KAGRA \cite{Thrane:2013npa,Thrane:2014yza,Coughlin:2016vor,Himemoto:2017gnw,Coughlin:2018tjc,Himemoto:2019iwd,Meyers:2020qrb} as well as for the Einstein telescope (ET)~\cite{PhysRevD.104.122006}.
The difficulty in conducting a search for a GWB with colocated detectors was displayed by LIGO with its H1 and H2 detectors in its fifth science run; correlated noise prevented the search for an isotropic GWB for frequencies below 460 Hz~\cite{LIGOScientific:2014sej}.

The Einstein telescope is the European proposal for a Third-generation Earth-based interferometric GW detector \cite{Punturo:2010zz}. The ET is proposed to be a made up of six interferometers with opening angle of $\pi/3$ and arm lengths of 10 km, arranged in an equilateral triangle. In this paper, we ignore the details of the xylophone configuration~\cite{Hild:2010id} and treat ET as consisting of three interferometers; this will have no effect on our studies. Proposed locations for the ET are the Sos Enattos mine in Sardinia, Italy, and the Euregio Rhein-Maas at the intersection of the Belgian, Dutch, and German borders \cite{Amann:2020jgo}.

The effect of seismic and Newtonian noise (NN) on GW interferometers and the possibility to apply (offline) noise subtraction has been studied extensively both for the second \cite{PhysRevD.86.102001,PhysRevD.92.022001,Coughlin_2014_seismic,Coughlin_2016,PhysRevLett.121.221104,Tringali_2019,Badaracco_2020} and third generation \cite{Badaracco_2019,NN_Sardinia2020,10.1785/0220200186,Bader_2022,Koley_2022} interferometric GW detectors. Ambient seismic fields rapidly lose coherence over large distances at the frequency range of interest of Earth-based GW interferometers (above several Hz) \cite{Coughlin_2014_seismic,Coughlin2019_seismic}. However, since the three interferometers of the ET triangular configuration will be (nearly) co-located, it is interesting to have a more detailed study of correlated seismic noise, the resulting NN and their impact on the ET.

To a first extent, one can typically assume that seismic fields over the scale of 10 km will be no longer correlated above a couple of Hertz \cite{Coughlin2019_seismic} In future studies, one might need to test this hypothesis in the case of an underground environment, preferable including the detector infrastructure. However, the end mirror of one of the ET-interferometers will be at a distance of several hundreds of meters from the input mirror of another ET-interferometer, e.g. 300\,m to 500\,m \cite{ETdesignRep}. In this paper we will expand on previous studies of seismic correlated noise \cite{Coughlin_2014_seismic,Coughlin2019_seismic} where we will focus on distance scales of 200\,m to 810\,m using both surface and underground sensors in a frequency range of 0.05\,Hz - 50\,Hz. Furthermore, we will use these seismic correlations to discuss the amplitude of correlated NN on a length scale of several hundreds of meters.
Please note that this paper does not contain a site comparison and the seismic spectra that will be used were selected based on the grid spacing of the installed sensors. The statements on the impact of correlated seismic and NN fields on the ET, will be of general nature, regardless of the exact location of the ET.

In Sec. \ref{sec:CorrSeisNoise} we investigate correlated seismic fields, which will be used in Sec. \ref{sec:CorrNN} to compute correlated NN. In Sec. \ref{sec:NoiseProjection} we introduce the formalism used to understand the impact of correlated noise sources on the search for an isotropic GWB. In Sec. \ref{sec:Results} we present upper limits on the maximal allowed seismic coupling such that correlated seismic noise does not affect the search for a GWB. Also the impact of correlated NN on the ET is discussed. In Sec. \ref{sec:Conclusion} we conclude our results and present an outlook for future work.

\section{Correlated seismic noise}
\label{sec:CorrSeisNoise}

To study the correlations of seismic noise at distances between 200\,m and 810\,m we use two different sensor networks. The first is an array of surface geophones measuring the vertical seismic velocity, which was deployed near Terziet (Netherlands) \cite{3T_network}. This array covers (a part of) the region of the ET candidate site at the Euregio Rhein-Maas. The advantage of this network is its large scale, and the presence of many sensors with horizontal separation of interest to us. However, there are also several downsides, namely the large level of anthropogenic activity in the region, the absence of an extensive underground network and the limited operation time of the network ($\propto$ month(s)). We also use underground sensors deployed in the former Homestake mine (USA) \cite{10.1785/0220170228}. Although the amount of sensors is more modest compared to the Terziet array, there are underground seismic measurements at several depths as well as horizontal separations of interest to our study. Furthermore, the Homestake mine is a seismically quiet environment and data are available for almost two years.

\begin{table*}[]
\begin{tabular}{|l|l|l|l|l|l|l|}
\hline
Location & Sensor Type & Name sensor 1 & Names sensor 2 & Horizontal distance & Depth & Direction \\ \hline
Terziet & Geophone & YKNVA & YONYA & $\sim 200 $ m & 0 m & Vertical \\ \hline
Terziet & Geophone & YCQGA & YIQEA & $\sim 300 $ m & 0 m & Vertical \\ \hline
Terziet & Geophone & XPPNA & XIPOA & $\sim 400 $ m & 0 m & Vertical \\ \hline
Terziet & Geophone & XPPNA & XYPWA & $\sim 500 $ m & 0 m & Vertical \\ \hline
Terziet & Geophone & YLOWA & YCPBA & $\sim 600 $ m & 0 m & Vertical \\ \hline
Terziet & Geophone & YKNVA & YSOJA & $\sim 700 $ m & 0 m & Vertical \\ \hline
Homestake & Seismometer & A2000 & B2000 & $\sim 255$ m & 610 m & 3-axial \\ \hline
Homestake & Seismometer & D2000 & E2000 & $\sim 405$ m & 610 m & 3-axial \\ \hline
Homestake & Seismometer & A2000 & D2000 & $\sim 580$ m & 610 m & 3-axial \\ \hline
Homestake & Seismometer & B2000 & D2000 & $\sim 810$ m & 610 m & 3-axial \\ \hline
\end{tabular}
\caption{Table summarizing the sensor pairs that are used in the correlation analysis in this paper. Please note that a depth of 0 m implies the sensor is located at the surface, however this does not imply the different sensors are at the same height above sea-level. The given distance is accurate up to 5 m.}
\label{tab:sensors}
\end{table*}

\hspace{1cm}
\paragraph*{Surface data from the Terziet geophone array} \hspace{1cm} \\

To characterize the seismic environment at the ET candidate site Euregio Rhein-Maas, hundreds of sensors (geophones, seismometers, ...) where deployed at the beginning of 2020 \cite{3T_network}. From this network we use a handful of geophones which where deployed in the region from one to several months. The sensor pairs used in the analysis are summarized in Tab. \ref{tab:sensors}. The sensor pairs were solely selected on their horizontal separation and no further investigations were performed concerning their geological location and/or their seismic environment. However another study \cite{SlidesSoumen} has shown the observed seismic spectrum vary significantly depending on the location of the sensors. It shows that the sensors which form our 300 m, 400 m and 500 m are intrinsically more quiet compared to the other sensors used in our analysis.

As mentioned earlier, the distance between the central station of one interferometer at the ET and the terminal station of another interferometer is proposed to be around 300\,m-500\,m. Therefore we decided to present coherence and cross-power spectral density (CSD) results for the data taken from the geophones with a horizontal separation of 400\,m ('XPPNA-XIPOA'-pair). Other distances will also be presented and compared with the 400\,m pair.

\begin{figure}
    \centering
    \includegraphics[width=\linewidth]{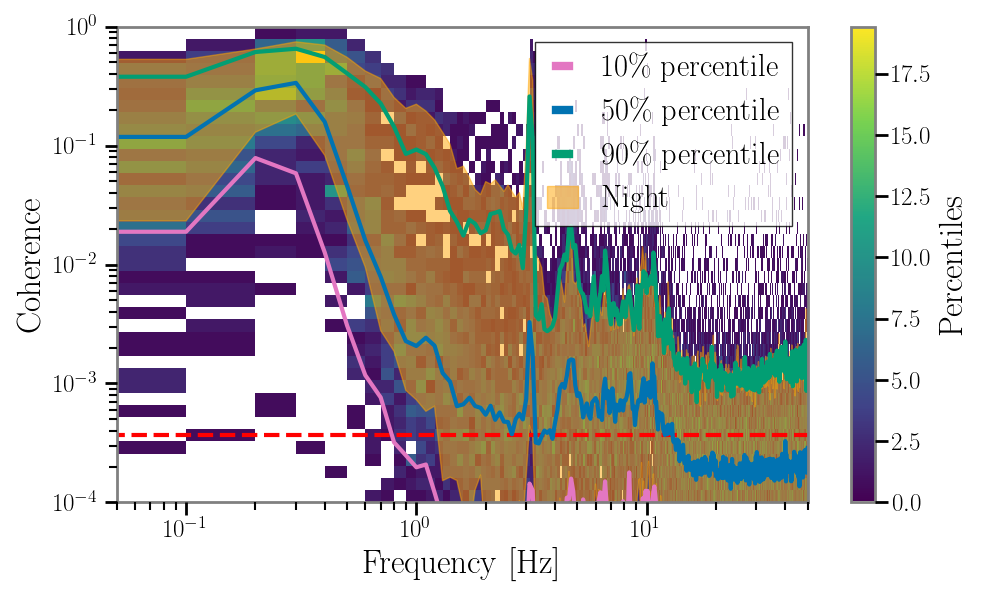}
    \caption{The coherence between the surface geophones XPPNA and XIPOA with an approximate distance of 400\,m. The data (10:15:01 UTC 07 Nov. 2020 - 12:13:59 UTC 5 Dec. 2020) are analysed using 10 second long segments which are averaged per 4h-window. The 10$^{\text{th}}$, 50$^{\text{th}}$ and 90$^{\text{th}}$ percentiles are shown in pink, blue and green, respectively. The percentiles as well as the counts are based on combined day and night data whereas the orange band represents the 10$^{\text{th}}$, to 90$^{\text{th}}$ percentiles during nighttime. The red dashed line represents the level of coherence expected from Gaussian data which goes approximately as 1/N, where N is the number of time segments over which was averaged. }
    \label{fig:Terziet400m_Coh}
\end{figure}

\begin{figure}
    \centering
    \includegraphics[width=\linewidth]{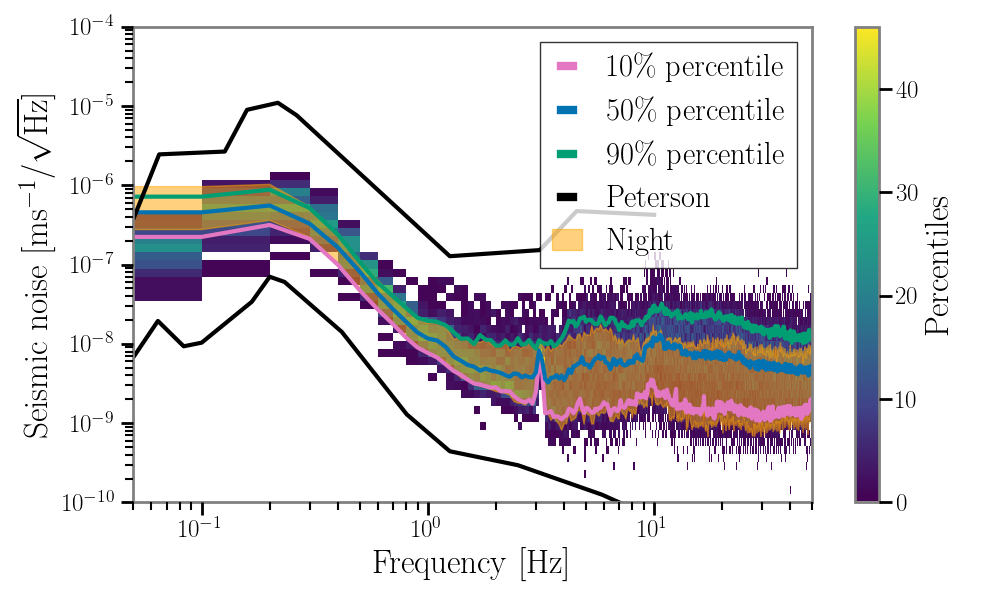}
    \caption{The CSD of the surface geophones XPPNA and XIPOA with an approximate distance of 400\,m. The data (10:15:01 UTC 07 Nov. 2020 - 12:13:59 UTC 5 Dec. 2020) are analysed using 10 second long segments which are averaged per 4h-window. The 10$^{\text{th}}$, 50$^{\text{th}}$ and 90$^{\text{th}}$ percentiles are shown in pink, blue and green, respectively. The percentiles as well as the counts are based on combined day and night data whereas the orange band represents the 10$^{\text{th}}$, to 90$^{\text{th}}$ percentiles during nighttime. The black curves represent the low and high noise models by Peterson \cite{Pet1993}.}
    \label{fig:Terziet400m_CSD}
\end{figure}

Fig. \ref{fig:Terziet400m_Coh} represents the coherence for the vertical geophones, whereas Fig. \ref{fig:Terziet400m_CSD} shows the CSD of the seismic spectrum in units of speed. Given the limited data (10:15:01 UTC 07 Nov. 2020 - 12:13:59 UTC 5 Dec. 2020 for XPPNA-XIPOA) we use a frequency resolution of 0.1Hz after which we average the data over stretches of 4 hr. This implies we are unable to get detailed results of the microseism peak near 0.2Hz. However, in this paper we are mainly interested in frequencies above 1\,Hz since the ET will be insensitive to GWs with lower frequencies.
Furthermore, we also show the 10\% and 90\% percentiles of the coherence as well as CSDs, measured during the local night time, represented as an orange band. For the local night time we have chosen 21:15 to 7:15 local time\footnote{We start 15 minutes after the hour rather than at the top of the hour since, the sensor started operating at 10:15 UTC and we analyse the data in stretches of one hour.}.

Based on Fig. \ref{fig:Terziet400m_Coh} we state that during 50\% of the time there is significant \footnote{We consider the coherence to be significant if the coherence is greater than 1/N, where N is the number of time segments over which was averaged. This 1/N is namely the approximate level of coherence expected from Gaussian data. } coherence up to $\sim$ 10\,Hz and at least 10\% of the time the coherence is significant up to 50\,Hz. The Seismic spectrum observed at these frequencies are approximately $10^{-9} \text{ ms}^{-1}/\sqrt{\text{Hz}}$ to $10^{-8}\text{ ms}^{-1}/\sqrt{\text{Hz}}$.

The local night time should give an indication of a quiet time when the effect of anthropoghenic noise is lower. Whereas the levels of observed coherence are comparable or marginally lower during day and night, the observed CSD is (marginally) lower during the night. This behaviour can be expected since during the night-time fewer local anthropogenic sources might disturb the coherence from the typically smaller ambient seismic fields.
We note that for some other pairs of geophones the difference was more pronounced. 

Around 0.2\,Hz we clearly observe the microseism peak and the observed cross-correlation spectrum lies within the seismic low noise and high noise models of Peterson \cite{Pet1993}. Note however, that these noise models were constructed for the amplitude/power spectral density of one geophone and not the cross spectral density of two geophones separated by several hundreds of meters. 

\begin{figure}
    \centering
    \includegraphics[width=\linewidth]{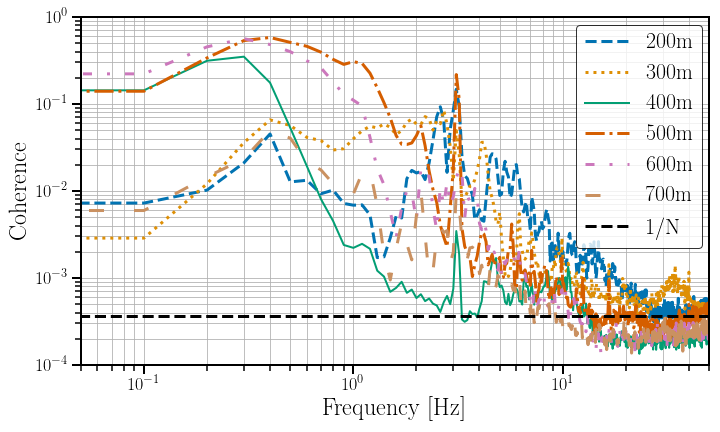}
    \caption{The median coherence of the surface geophones as a function of distance. The data are analysed using 10 second long segments which are averaged per 4h-window. Please note that the data at different distances is taken during different times as well as varying period (22 days to 28 days).}
    \label{fig:Terziet_dist_Coh}
\end{figure}

\begin{figure}
    \centering
    \includegraphics[width=\linewidth]{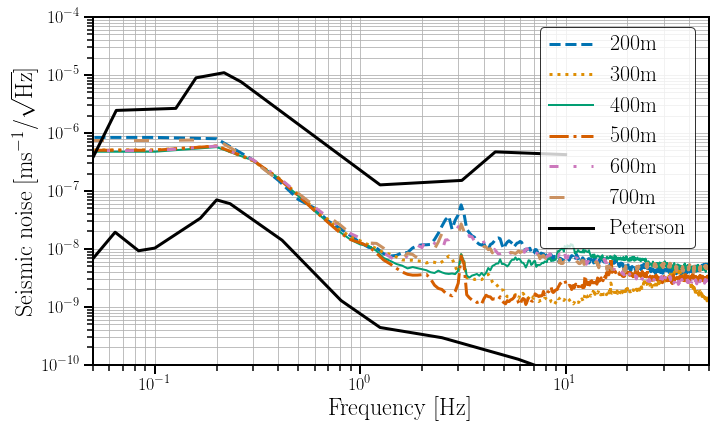}
    \caption{The median CSD of the surface geophones as a function of distance. The data are analysed using 10 second long segments which are averaged per 4h-window. Please note that the data at different distances is taken during different times as well as varying period (22 days to 28 days). As a comparison, the Peterson low and high noise models are shown in black.}
    \label{fig:Terziet_dist_CSD}
\end{figure}

In Fig. \ref{fig:Terziet_dist_Coh} and Fig. \ref{fig:Terziet_dist_CSD} we present the 50\% percentile of the coherences and of the CSDs measured for different geophone separations, respectively.
Fig. \ref{fig:Terziet_dist_Coh} shows that the seismic coherence for distances between 200\,m and 700\,m is significant up to $\sim$ 10\,Hz for 50\% of the time. Both Fig. \ref{fig:Terziet_dist_Coh} and Fig. \ref{fig:Terziet_dist_CSD} show that there is no clear relation between the observed coherences (or the CSDs) and the horizontal separations between the geophones. If the seismic field were perfectly isotropic and homogeneous, we should have observed high coherence up to around 0.1\,Hz with a faster decrease for larger geophone separations (Equation (1) of \cite{Yokoi2008}). We do not see this behaviour in the data, on the contrary, we observe more coherence for separations of 400, 500, 600\,m and reduced coherence for 200, 300 and 700\,m. A possible explanation can be the large amount of anthropogenic noise sources in the region. This can lead to anisotropies in the seismic field. Furthermore, the geophone pairs are located in positions with different levels of ambient seismic fields \cite{SlidesSoumen}. Also, the directions of the seismic waves can affect the coherences. The aim of this paper is to get an order of magnitude estimate of the correlated seismic spectrum between sensors separated by several hundreds of meter and use these to make an estimate of the subsequent NN and their effect on the search for a GWB. However, the variation of both the coherence and seismic spectrum (in units of speed) presented in respectively Fig. \ref{fig:Terziet_dist_Coh} and Fig \ref{fig:Terziet_dist_CSD} indicate more precise (site-specific) studies are needed to fully understand the effect of local geology and anisotropies in the seismic field.

\hspace{1cm}
\paragraph*{Underground data from the Homestake seismometer array} \hspace{1cm} \\

For studying the underground correlations we will use underground seismometers located at the 2000\,ft-level of the former Homestake mine \cite{10.1785/0220170228}. The ET is proposed to be located at a depth of around 200-300\,m underground \cite{ETdesignRep}, this is much less than our choice of 2000\,ft $\approx$ 610\,m. However, the choice of this depth was driven by the presence of multiple seismometers deployed with horizontal separations between $\sim$125\,m and $\sim$1200\,m. Moreover, at this depth, the effect of surface waves will be even more suppressed and the correlations will provide realistic insights regarding correlations of seismic body waves at horizontal distances of interest for ET.

The seismometer pairs of which we will present the results are introduced in Tab. \ref{tab:sensors}

Similar to what was done with the data of the Terziet array presented earlier, we will use data measured at a distance of 405\,m, that is between the 'D2000' and 'E2000' stations, as an example. Afterwards we will compare it with data from other distances.

\begin{figure}
    \centering
    \includegraphics[width=\linewidth]{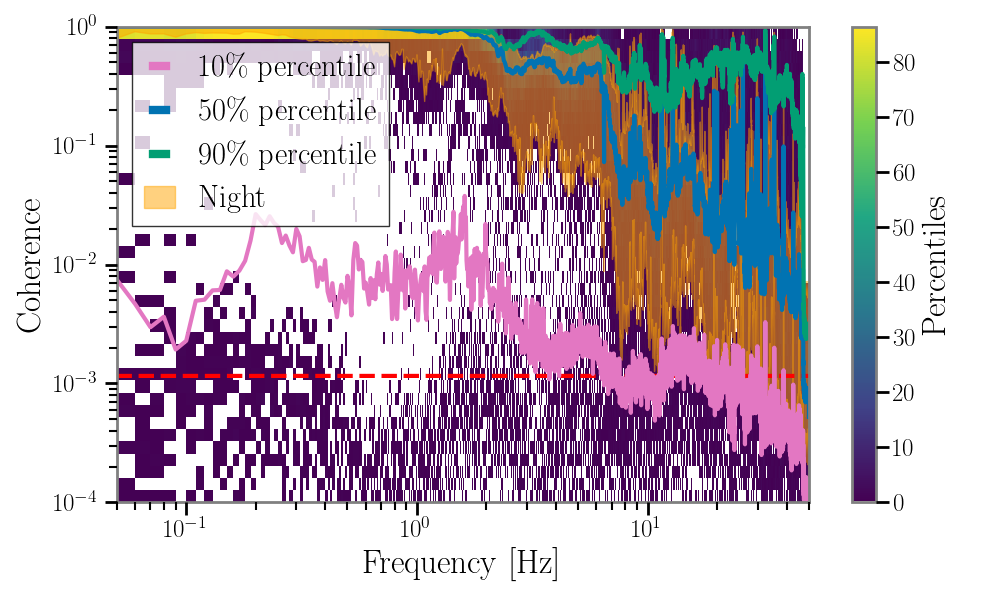}
    \caption{The coherence between the underground (depth $\approx$ 610m) seismometers (vertical component) D2000 and E2000 with an approximate horizontal distance of 405m. The data (March 2015 to Dec 2016) are analysed using 100 second long segments which are averaged per 24h-window. The 10$^{\text{th}}$, 50$^{\text{th}}$ and 90$^{\text{th}}$ percentiles are shown in respectively pink, blue and green. The percentiles as well as the counts are based on combined day and night data whereas the orange band represents the 10$^{\text{th}}$, to 90$^{\text{th}}$ percentiles during nighttime. The red dashed line represents the level of coherence expected from Gaussian data which goes approximately as 1/N, where N is the number of time segments over which was averaged.}
    \label{fig:D2000-E2000_Coh}
\end{figure}

\begin{figure}
    \centering
    \includegraphics[width=\linewidth]{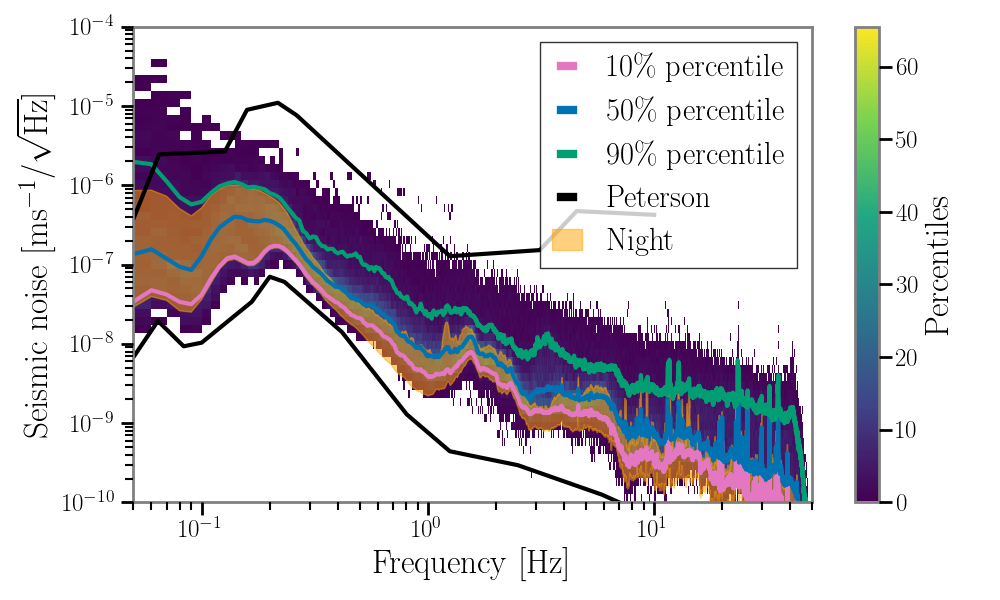}
    \caption{The CSD between the underground (depth $\approx$ 610m) seismometers (vertical component) D2000 and E2000 with an approximate horizontal separation of 405m. The data (March 2015 to Dec 2016) are analysed using 100 second long segments which are averaged per 24h-window. The 10$^{\text{th}}$, 50$^{\text{th}}$ and 90$^{\text{th}}$ percentiles are shown in respectively pink, blue and green. The percentiles as well as the counts are based on combined day and night data whereas the orange band represents the 10$^{\text{th}}$, to 90$^{\text{th}}$ percentiles during nighttime. The black curves represent the low and high noise models by Peterson \cite{Pet1993}.}
    \label{fig:D2000-E2000_CSD}
\end{figure}

Whereas the Terziet geophones were measuring only the vertical seismic velocity, the Homestake seismometers can measure it in the vertical as well as in the horizontal directions (called North-South and East-West). Fig. \ref{fig:D2000-E2000_Coh} represents the coherence for the vertical seismometers, while Fig. \ref{fig:D2000-E2000_CSD} shows the accompanying CSD. 
Similar to the results for Terziet, we show a percentile plot together with the 10$^{\text{th}}$, 50$^{\text{th}}$ and 90$^{\text{th}}$ percentiles. Given the larger amount of data available for Homestake we used a frequency resolution of 0.01\,Hz and averaged over 24\,h. The data represented here uses 600\,days recorded between March 2015 and December 2016.
Furthermore we also show the 10\% and 90\% percentiles of the coherence as well as CSDs, measured during the local night time, represented as an orange band. We define local night time between 04:00 UTC and 14:00 UTC. This matches 21:00-07:00 local time during the 'standard time' (that is during the winter period).

Above $\sim$ 40\,Hz the response of the seismometers in the set-up at the Homestake mine decreases rapidly and the results should not be trusted. Fig. \ref{fig:D2000-E2000_Coh} shows that 90\% of the time there is significant coherence up to 20\,Hz and more than 50\% of the days even up to 40\,Hz.
The seismic spectrum observed at these frequencies are approximately $10^{-10} \text{ ms}^{-1}/\sqrt{\text{Hz}}$ to $10^{-8}\text{ ms}^{-1}/\sqrt{\text{Hz}}$. We notice that, definitely at frequencies above a couple of Hz, the seismic CSD at Homestake is considerable lower than the CSDs observed at Terziet. 

Around the microseism peak ($\sim$0.2Hz) one typically expects high levels of coherence, which is not the case for the 10\% percentile in Fig. \ref{fig:D2000-E2000_Coh}. Further investigation of the data showed that some of the seismometers observed loud excess on certain days, likely linked to loud anthropogenic and close-by events. These specific anthropogenic and loud disturbances might be specific to Homestake but also at ET many local activities will take place. To this extent the results might give a realistic -- or at most conservative -- prediction of possible levels of correlated seismic noise at the ET. An ideal scenario is represented by the night time measurements which are not affected by these loud events.

The spread of the coherence and the CSD is lower during night times and above 10\,Hz the CSD is considerably lower compared to those measured during day and night combined (Fig \ref{fig:D2000-E2000_Coh} and \ref{fig:D2000-E2000_CSD}). This is what one might expect to observe since at these frequencies many noise sources are anthropogenic which are expected to be lower during local night time. Even though the CSD observed during the night time is considerably lower, the effect on the 10\% percentile is modest, i.e. a factor of $\sim$2-3. The 90\% percentile during the night is at a comparable level as the 50\% percentile measured during the day.

Similarly we investigated whether a reduction in correlated seismic noise could be observed during the weekends compared to the entire data-set. However, in this case no significant effect was observed, which led us to conclude local night times are the most quiet periods due to lower anthropogenic noise.

\begin{figure}
    \centering
    \includegraphics[width=\linewidth]{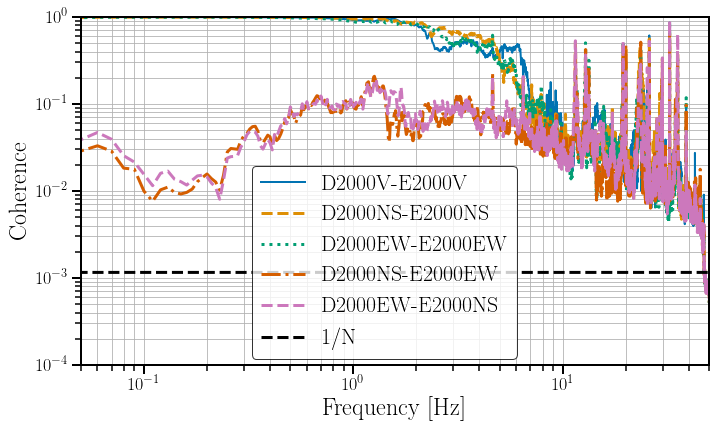}
    \caption{The median coherence of the underground (depth $\approx$ 610m) seismometers as a function of the different orientations, at a horizontal separation of 405m. The data (March 2015 to Dec 2016) are analysed using 100 second long segments which are averaged per 24h-window. The black dashed line represents the level of coherence expected from Gaussian data which goes approximately as 1/N, where N is the number of time segments over which was averaged.}
    \label{fig:Homestake_dir_Coh}
\end{figure}

\begin{figure}
    \centering
    \includegraphics[width=\linewidth]{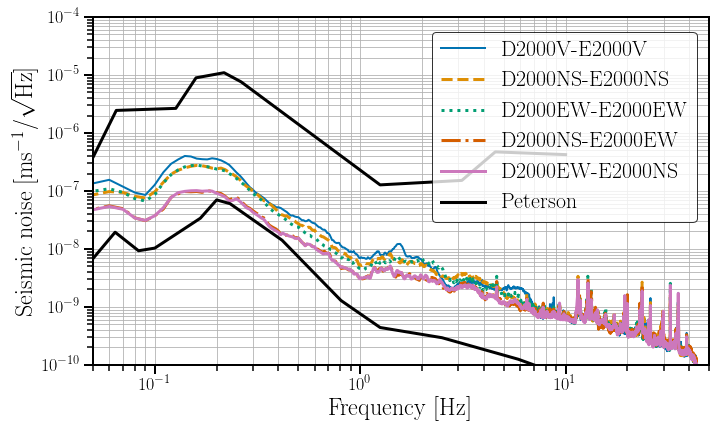}
    \caption{The median CSD of the underground (depth $\approx$ 610\,m) seismometers as a function of the different orientations, at a horizontal separation of 405m. The data (March 2015 to Dec 2016) are analysed using 100 second long segments which are averaged per 24h-window. As a comparison, the Peterson low and high noise models are shown in black.}
    \label{fig:Homestake_dir_CSD}
\end{figure}

In Fig.~\ref{fig:Homestake_dir_Coh} and Fig.~\ref{fig:Homestake_dir_CSD} we present the 50$^{\text{th}}$ percentile of the coherences, respectively of the CSDs measured for different seismometer channels (i.e. vertical and horizontal) and with a horizontal separation of $\sim$ 405\,m. At low frequencies the coherences between perpendicular channels (NS-EW and EW-NS) are lower than those between parallel channels; this is not true anymore above 6-7\,Hz, where the coherences start to be spoiled by anthropogenic noise, which we can expect containing many incoherent sources. The difference in CSD for perpendicular and parallel seismometers is negligible above 2-3Hz.

\begin{figure}
    \centering
    \includegraphics[width=\linewidth]{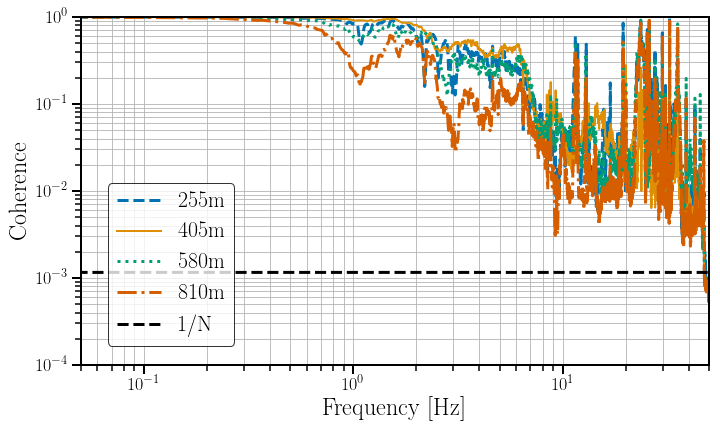}
    \caption{The median coherence of the underground (depth $\approx$ 610\,m) seismometers (vertical component) as a function of the horizontal distance. The data (March 2015 to Dec 2016) are analysed using 100 second long segments which are averaged per 24h-window. The black dashed line represents the level of coherence expected from Gaussian data which goes approximately as 1/N, where N is the number of time segments over which was averaged.}
    \label{fig:Homestake_dist_Coh}
\end{figure}

\begin{figure}
    \centering
    \includegraphics[width=\linewidth]{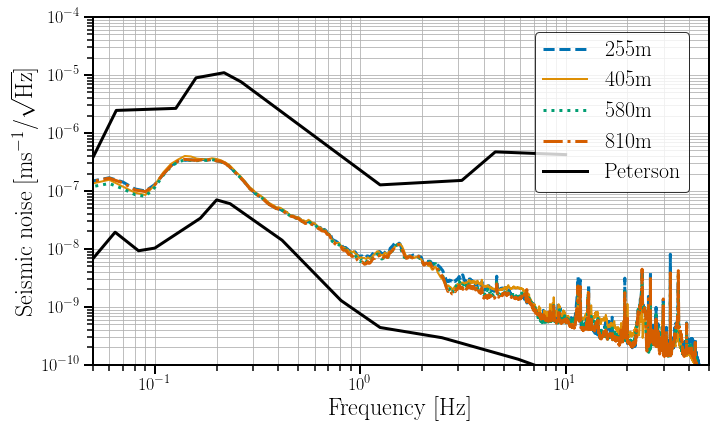}
    \caption{The median CSD of the underground (depth $\approx$ 610m) seismometers as a function of the horizontal distance. The data (March 2015 to Dec 2016) are analysed using 100 second long segments which are averaged per 24\,h-window. As a comparison, the Peterson low and high noise models are shown in black.}
    \label{fig:Homestake_dist_CSD}
\end{figure}

In Fig. \ref{fig:Homestake_dist_Coh} and Fig. \ref{fig:Homestake_dist_CSD} we present the 50$^{\text{th}}$ percentiles of the coherences, respectively the CSDs measured at different seismometers separations.
We can notice that the coherences diminish for larger seismometer separations, except for the 405\,m separation. This might be explained by a partial anisotropy in the seismic field which can lead to a higher apparent velocity and thus higher coherence. Furthermore, the CSDs are very similar regardless of the horizontal distance between the horizontal seismometers.

\section{Correlated Newtonian noise}
\label{sec:CorrNN}

Newtonian noise (NN) is a disturbance produced in GW detectors by local fluctuations in the gravitational field ~\cite{PhysRevD.30.732,2019,PhysRevD.58.122002}. Any change in the local density of rocks or air will in turn generate gravity fluctuations (it is a simple consequence of Newton's gravity law). These density variations are mainly generated by passing seismic waves ~\cite{PhysRevD.86.102001,Beker_2012} and atmospheric phenomena \cite{creighton2008tumbleweeds}: from here we can distinguish NN of seismic and atmospheric origin. In this paper we will only discuss the former one.

NN affects GW detectors by directly exerting a force on their test masses. This effect mainly affects the low-frequency region between $\sim$1-30\,Hz, where it will be the limiting noise source to the ET sensitivity \cite{Amann:2020jgo}. There are no easy ways to shield the test masses from NN ~\cite{Harms2014}, therefore it will have to be reduced by the implementation of (offline) noise cancellation ~\cite{Driggers2012}. 
To estimate the correlated NN of seismic origin we will consider the previously obtained results using seismic measurements of the Terziet surface array and the Homestake underground array.

Seismic waves can be divided in surface and body waves. Among surface waves, Rayleigh waves are the only ones producing density fluctuations and therefore NN~\cite{novotny1999seismic}.
Body waves, instead propagate in the underground bulk and can be divided into compression waves (P-waves) and shear waves (S-waves) ~\cite{novotny1999seismic, 2019}. Both P- and S-waves contribute to the NN noise from body waves.\\
The aim of this study is to provide an estimate of the correlated NN at a horizontal distance of $\sim$ 400\,m. We will assume a homogeneous and isotropic bulk as well as a flat surface topology. 

\subsection{Newtonian noise formalism}

\subsubsection{NN from Rayleigh waves}
Rayleigh waves produce NN through surface displacement, cavern walls displacement and rock compression. We need to take all these effects into account if we want to estimate NN from Rayleigh waves. The NN strain spectral density is a function of the depth, $h$, of the GW detector, which we can write as~\cite{Amann:2020jgo}:
\begin{equation}\label{eq:Rayleigh}
\begin{aligned}
    S_{h,\text{Rayleigh}}(f) =  &\left( \sqrt{2} \pi G\gamma\rho_{0,\text{Surface}}\right)^2 \frac{1}{L^2 (2\pi f)^4} \\
    & \mathcal{R}(h,f) S_{\xi_z}(f)
\end{aligned}
\end{equation}
with $S_{\xi_z}$ the PSD, or in our case the CSD, of the vertical displacement of the Rayleigh wave and $\rho_{0,\text{Surface}} = 2 \ 800 \text{kg\,m}^{-3}$ the surface density \cite{harms2022lower}. $\gamma$ is a factor given by the elastic properties of the half space that takes into account the partial cancellation of the NN given by the compression/decompression of the rocks caused by the Rayleigh waves below the surface: we assume $\gamma = 0.8$ (see Fig 10 of \cite{2019}). $\mathcal{R}(h,f)$ is given by,
\begin{equation}
    \mathcal{R}(h,f) =  \left| \frac{-k_R(1+\zeta)e^{-k_Rh} + \frac{2}{3}\left( 2k_Re^{-q_Ph} + \zeta q_Se^{-q_Sh}\right)}{k_R(1-\zeta)} \right|^2
\end{equation}
Where for sake of compactness we did not explicitly write the frequency dependence of $k_R,\,q_P,\,q_S$ and $\zeta$. Moreover here: $k_R = 2 \pi f/v_R(f)$ is the Rayleigh wave number, $h$ the detector's depth and: 
\begin{align}
&q_P(f) =  \frac{2\pi f}{v_R(f)v_P}\sqrt{v_P^2 - v_R^2(f)}\\
&q_S(f) =  \frac{2\pi f}{v_R(f)v_S}\sqrt{v_S^2 - v_R^2(f)}\\
&\zeta(f) = \sqrt{\frac{q_P(f)}{q_S(f)}}
\end{align}
 Additionally we assumed a Rayleigh wave velocity given by $v_R(f) = 2000 \text{ m\,s}^{-1}e^{-f/4\text{Hz}} + 300\text{ m\,s}^{-1}$ in the frequency range 1\,Hz - 100\,Hz~\cite{BoEA2002ch2}. For the velocity of P and S waves we use the values given by the estimate of Bader \textit{et al.}~\cite{Bader_2022}, since we will be using the surface measurements of Terziet for the calculation of the NN from Rayleigh waves. We use $v_P(f) = 4.05 \text{ km.s}^{-1}$ and $v_S(f) = 2.4 \text{ km.s}^{-1}$ estimated for a depth $h>58.2$~\cite{Bader_2022}. 

\begin{figure}
    \centering
    \includegraphics[width=\linewidth]{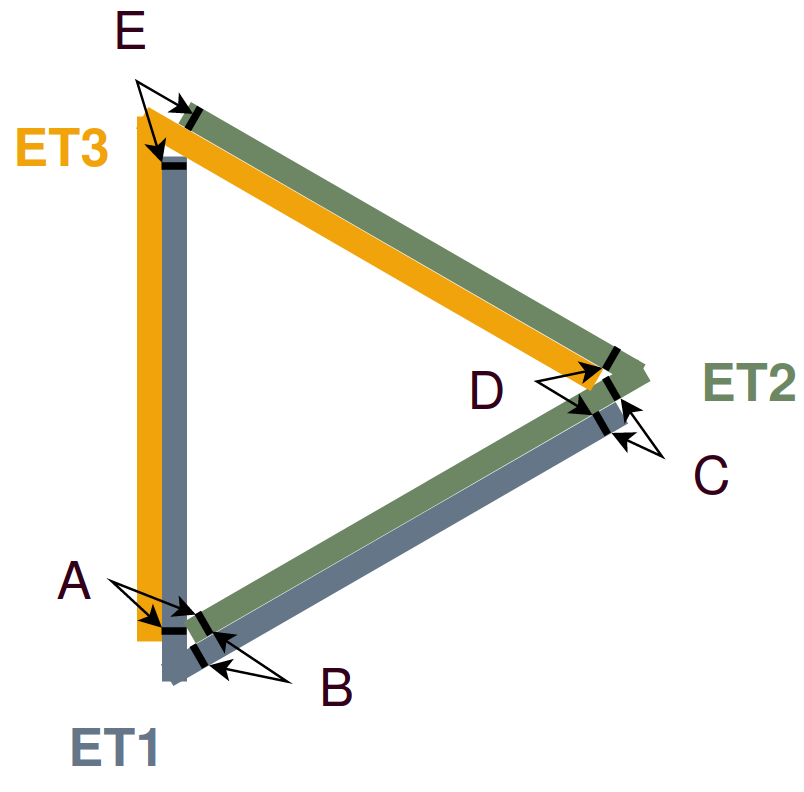}
    \caption{Scheme of the ET configuration (the low- and the high-frequency detectors are not showed). Considering the ET1-ET2-baseline, we can identify 5 possible coupling locations where the NN can correlate: A to E. B and C involve the coupling of two aligned mirrors, while A,D and E involve mirrors which are rotated by 60\textdegree with respect to each other.}
    \label{fig:ETDiagram}
\end{figure}

The term $1/L^2 (2\pi f)^4$ accounts for the conversion factor from acceleration to strain, where one test mass was considered.  
In this study instead, we are interested in the effect of correlated noise on the baseline formed by two interferometers, e.g. ET1 and ET2. Fig. \ref{fig:ETDiagram} illustrates that there are five possible locations where correlated noise can be introduced in the ET1-ET2-baseline. The horizontal separation for aligned mirror pairs (B and C) is about 300m - 500m, whereas one can calculate the distance for the other pairs (A, D and E) to be about 330m - 560m \cite{ETdesignRep}. In this paper we will multiply this NN contribution by a factor of 5, assuming conservatively the incoherent sum of the correlated NN between all the mirror pairs. In reality, some of them could show correlated behaviour which would lead to a factor lower than 5. Please also note that some pairs involve aligned mirror pairs whereas others are misaligned by 60\textdegree.
To estimate the NN from Rayleigh waves we used the analytical model taken from \cite{Amann:2020jgo} rather than performing a numerical analysis as done in \cite{Bader_2022}. This approach is site independent and gives an idea on the orders of magnitude involved. A site-specific study could
be envisioned in the future and will yield more accurate results. This is however beyond the scope of this paper.

\subsubsection{NN from body-waves}

Seismic waves from body waves produce NN through two mechanisms: displacement of cavity walls (both P and S-waves) and compression/decompression of the bulk. Taking these two mechanisms into account we can get an analytical estimate for the body waves NN \cite{Amann:2020jgo}:

\begin{equation}\label{eq:BWNN}
    S_{\text{Body-wave}}(f) = \left(\frac{4\pi}{3}G\rho_{0,\text{Bulk}} \right)^2 (3p+1) \frac{1}{L^2(2\pi f)^4}S_{\xi_x}(f)
\end{equation}
where $S_{\xi_x}$ represents the PSD, or in our case the CSD, of the displacement caused by the body-waves along the arm direction.
Eq. \ref{eq:BWNN} assumes again the contribution of one test mass. In our results we will, as explained earlier, multiply by a factor of 5, taking into account the five different coupling locations for correlated noise, indicated by letters A-E in Fig. \ref{fig:ETDiagram}. The measurements at Homestake indicate seismic correlation measured with sensors along parallel or perpendicular directions are at the same level above 2-3\,Hz, which is the main region of interest.
The density $\rho_{0,\text{Bulk}} = 2800 \text{kg\,m}^{-3}$ is the bulk density. Moreover, the parameter $p$ accounts for the different mixing ratio of P- and S-waves. Given that we are performing a site-independent study, we can assume $p$ to be 1/3. This accounts for an equal energy distribution between P-waves (with one polarization) and S-waves (two possible polarizations) \cite{2019}.

\subsection{Result of the NN projection in ET band}
To have an estimate of the NN correlations that might affect stochastic searches in the ET, we need to calculate the NN strain for the correlated Terziet data (Rayleigh waves case, see Eq.~\ref{eq:Rayleigh}) and for the correlated Homestakes data (body waves case, see Equation~\ref{eq:BWNN}). We will compare the NN estimates with the ET design sensitivity, for which we use the ET-Xylophone design also known as 'ET-D' \cite{Hild_2009,Hild:2010id}.

\subsubsection{NN from Rayleigh waves}

To characterize the correlated NN produced by Rayleigh surface waves in ET, we use the CSD of the Terziet geophone array XPPNA and XIPOA presented in  Fig.~\ref{fig:Terziet400m_Coh}, we are therefore considering a horizontal separation of $\sim$ 400\,m. 
To compute the Rayleigh NN, we use Eq.~\ref{eq:Rayleigh} with the 10$^{\text{th}}$, 50$^{\text{th}}$ and 90$^{\text{th}}$ percentile of the CSD, using both day and night. 

\begin{figure}
    \centering
    \includegraphics[width=\linewidth]{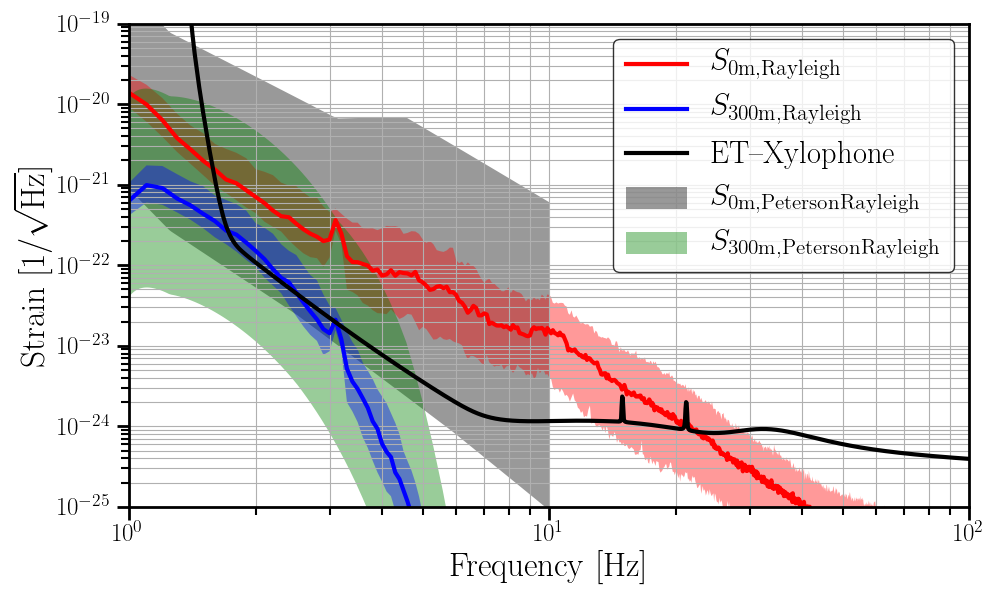}
    \caption{Strain of the NN with CSD of the surface geophone XPPNA and XIPOA (see Fig.~\ref{fig:Terziet400m_CSD}) with a horizontal distance of 400\,m at the surface (red curve) and at 300 m depth (blue curve). In the two cases considered, the solid line corresponds to the NN strain from the 50 $\%$ percentile, whereas the associated surface is delimited by the 10$^{\text{th}}$ and 90$^{\text{th}}$ percentiles of the CSD. The respective gray and green surface, delimited by the low and high limits of Peterson measurement \cite{Pet1993}, are the Rayleigh NN strain of the surface and 300\,m depth. The black line is the ET-Xylophone design sensitivity.   }
    \label{fig:NNTerziet400mAll}
\end{figure}

In Fig.~\ref{fig:NNTerziet400mAll}, the red and blue lines show the correlated NN strain from the 50$^{\text{th}}$ percentile for different depths. The red line is calculated at the surface (${\rm depth} = 0{\rm \ m}$) and the blue line at a depth of 300\,m, (consistent with the future depth of ET~\cite{ETdesignRep}). We also evaluate the two limiting cases given by the 10$^{\text{th}}$ and 90$^{\text{th}}$ percentile of the CSD (red  and blue shaded areas in Fig~\ref{fig:NNTerziet400mAll} for 0 and 300\,m depth). 
About the blue curve we need some care. Indeed, in a homogeneous geology, Rayleigh waves speed would be related to the speed of shear waves in the following way: $v_R = 0.9 v_S$. If the geology is instead layered, the Rayleigh wave speed will become dispersive, i.e. dependent from the frequency \cite{novotny1999seismic}. In the latter case, we would also have different values of the P- and S-waves speeds in each layer. Concerning the body wave NN, we have that the S- and P-waves speed does not affect the model, since in the small cavern approximation it does not enter into play (while the density of the layers enters linearly in the model, which is therefore less sensitive to its variations along the layers). See Equation \ref{eq:BWNN}.
The Rayeligh NN model is instead different. Here the speed (of Rayleigh waves, as well as of P- and S-waves) enters exponentially in the model (Equation \ref{eq:Rayleigh}), which is therefore more affected by it. So, we must be careful in the speed modelling.
We evaluated three cases: the first, where we kept $v_S$ and $v_P$ constant, with values taken from \cite{Bader_2022} and with $v_R$ modelled as in \cite{Amann:2020jgo}. In the second and third case, instead, we modelled $v_P$ and $v_S$ to be frequency dependent in the following way: $v_P(f) = 2v_R(f)$ and $v_S(f) = 1.1v_R(f)$. Here $v_R(f)$ was modelled as in \cite{Amann:2020jgo} or taken from the Terziet measurements \cite{Bader_2022}. We found that the first model led to a slightly more conservative Rayleigh NN contribution, which is the reason why we used that.

This estimation can also be compared with the NN estimate calculated with Peterson's noise models (low and high)~\cite{Pet1993}. These are presented respectively in grey and green for the depths of 0 and 300\,m. We should note that the Peterson NN estimates represent a conservative NN correlation estimate, given that they represent the lower and higher limits for the power spectral density of seismic noise, therefore it will always be higher than the CSD. It is the same as having 100\% coherence. 

We compare the ET-Xylophone design sensitivity (black line) and the NN Rayleigh calculation for Terziet NN 90$^{\text{th}}$ percentile estimation. A detector located at the surface would therefore have a correlated NN above the ET-Xylophone design sensitivity from before 2\,Hz to 30\,Hz. At a depth of 300\,m,instead, the 90$^{\text{th}}$ percentile of CSD NN lies above the ET-Xylophone design sensitivity only up to around 3\,Hz. Again, we can easily see that building the ET underground can lead to a reduction of the Rayleigh NN. 

\subsubsection{NN from body-waves}

To characterize the correlated body waves NN for the ET, we use the CSD of the Homestakes underground seismometers, D2000 and E2000, in Fig.~\ref{fig:D2000-E2000_CSD}. We took a CSD$_{\rm tot}$ given by the mean of the two directions NS-NS and EW-EW. 
According to Fig.~\ref{fig:Homestake_dist_CSD}, we have no significant evidence of a difference in the CSDs for seismometers with distance separations between 225\,m and 810\,m. For the calculation of body wave induced NN, we use the same distance as the Rayleigh NN measurement (405\,m, from the Homestake data). 

\begin{figure}
    \centering
    \includegraphics[width=\linewidth]{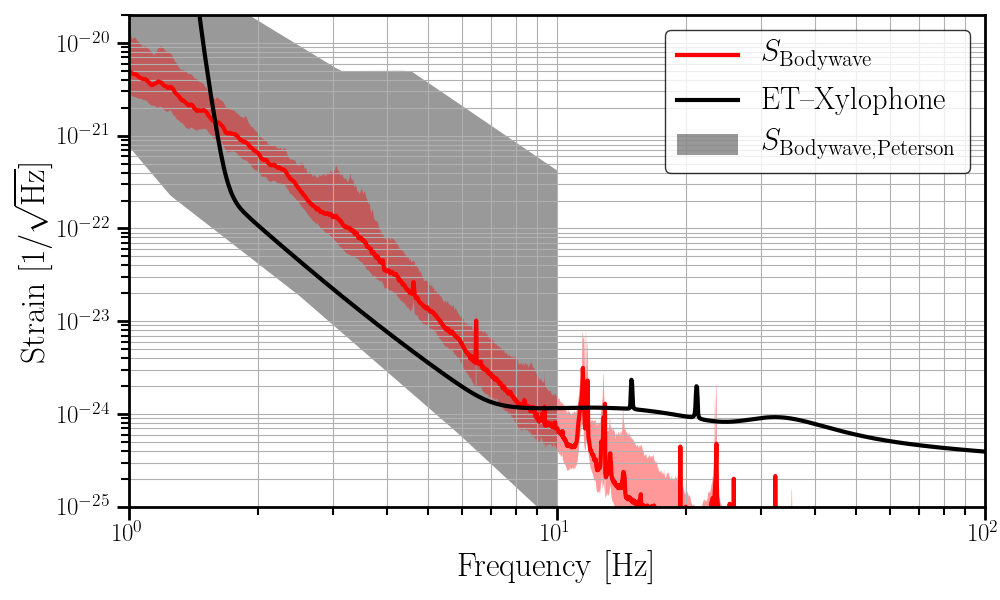}
    \caption{Strain of the NN with CSD of the Homestakes underground seismometers D2000 and E2000 vertical displacement measurement (see Fig.~\ref{fig:Homestake_dir_CSD}) with a horizontal distance of 405m at a depth of $\sim$ 610\,m (red curve). The solid line is the body waves NN strain from the 50 $\%$ percentile and the surface associated is delimited by the 10$^{\text{th}}$ and 90$^{\text{th}}$ percentiles CSD. The gray surface, delimited by the low and high limits of Peterson measurement \cite{Pet1993}, are the body wave NN strain at 610 m depth. The black line is the ET-Xylophone design sensitivity  }
    \label{fig:NNHomestake400mAll}
\end{figure}

To compute the body waves NN, we use Eq.~\ref{eq:BWNN} with the 10$^{\text{th}}$, 50$^{\text{th}}$ and 90$^{\text{th}}$ percentiles of CSD$_{\rm tot}$. In Fig.~\ref{fig:NNHomestake400mAll}, the red line is the correlated NN strain from the 50$^{\text{th}}$ percentile of the CSD. We also calculate the two limiting cases given by the 10$^{\text{th}}$ and 90$^{\text{th}}$ percentiles of the CSD (red surface in Fig~\ref{fig:NNHomestake400mAll}).  Again, Peterson's limits (low and high)~\cite{Pet1993} are used for comparison (grey surface in fig~\ref{fig:NNHomestake400mAll}). 

The NN body wave estimate for the 90$^{\text{th}}$ percentile CSD, exceeds the ET sensitivity curve starting from before 2\,Hz up to 10\,Hz. 

\section{Noise projection - formalism}
\label{sec:NoiseProjection}

Since the search for a GWB is very sensitive, if not the most, to correlated noise sources we will investigate the impact of the correlated seismic and Newtonian noise described in Sec. \ref{sec:CorrSeisNoise}, respectively Sec. \ref{sec:CorrNN}. In this section we will describe the relevant formalism to present and discuss the obtained results in Sec. \ref{sec:Results}.

In this paper we will only focus on the search for an isotropic GWB, of which one typically tries to measure its energy density, $\text{d}\rho_{\rm GW}$, contained in a logarithmic frequency interval, $\text{d}\ln f$. Furthermore one divides by the critical energy density $\rho_{\rm c} = 3H_0^2c^2/(8\pi G)$ for a flat Universe to construct a dimensionless figure of merit $\Omega_{\rm GW}(f)$~\cite{Christensen_2018,PhysRevD.46.5250,PhysRevD.59.102001,LivingRevRelativ20}:

\begin{equation}
    \label{eq:omgeaGWB}
    \Omega_{\rm GW}(f) = \frac{1}{\rho_{\rm c}}\frac{\text{d}\rho_{\rm GW}}{\text{d}\ln f}~,
\end{equation} 
where $H_0$ is the Hubble-Lemaître constant, $c$ is the speed of light and $G$ is Newton's constant. We use the 15-year Planck value of 67.9 km s$^{-1}$ Mpc$^{-1}$ for $H_0$~\cite{Planck:2015fie}. 

When searching for an isotropic, Gaussian, stationary and unpolarized GWB, one can construct the cross-correlation statistic $\hat{C}_{IJ}(f)$,
\begin{equation}
    \label{eq:cross-correlationstatistic}
\hat{C}_{IJ}(f) = \frac{2}{T_{\textrm{obs}}} \frac{{\rm{Re}}[\Tilde{s}^*_I(f)\Tilde{s}_J(f)]}{\gamma_{IJ}(f)S_0(f)}~,
\end{equation}
which is an unbiased estimator of $\Omega_{\rm GW}(f)$ in the absence of correlated noise~\cite{PhysRevD.59.102001,LivingRevRelativ20}. $I$ and $J$ represent the two interferometers and $\Tilde{s}_I(f)$ is the Fourier transform of the time domain strain data $s_I(t)$ measured by interferometer $I$. $\gamma_{IJ}$ is the normalized overlap reduction function which encodes the baseline's geometry~\cite{PhysRevD.46.5250,Romano:2016dpx}. $S_0(f)$ is a normalisation factor given by $S_0(f)=(9H_0^2)/(40\pi^2f^3)$ and $T_{\textrm{obs}}$ is the total observation time of the data-collecting period\footnote{The normalisation factor $S_0(f)$ for ET differs from that one of e.g. LIGO-Virgo-KAGRA by a factor of 3/4, due to the different opening angle between the interferometers' arms ($\pi$/2 for LIGO-Virgo-KAGRA and $\pi$/3 for ET) \cite{Romano:2016dpx}.}.

In line with an earlier study on the impact of correlated magnetic noise on the ET \cite{PhysRevD.104.122006} we will refer to the three different ET interferometers as ${\rm ET}_1, {\rm ET}_2, {\rm ET}_3$, which we assume to have identical sensitivity. Furthermore we neglect the difference in $\gamma_{IJ}$ between the baseline pairs $IJ =  {\rm ET}_1{\rm ET}_2; {\rm ET}_1{\rm ET}_3; {\rm ET}_2{\rm ET}_3$, since the relative difference between the overlap reduction functions of the different arms is smaller than $5\times 10^{-7}$ for frequencies under 1 kHz ~\cite{PhysRevD.104.122006}. In the remainder of the paper we will use the ${\rm ET}_1{\rm ET}_2$-baseline as our default observing baseline.

Similar to the magnetic cross-correlation described in earlier work~\cite{PhysRevD.104.122006,Thrane:2013npa,Thrane:2014yza}, we can construct equivalent cross-correlation statistics for the correlated seismic noise and NN:

\begin{equation}
    \label{eq:C_Seis_NN}
    \begin{aligned}
        \hat{C}_{{\rm seismic},{\rm ET}_1{\rm ET}_2}(f) =& \\
        |\Kappa_{\rm seismic,vth, ET}(f)|^2 &N_{{\rm seismic,v, ET}_1{\rm ET}_2} \\
        + |\Kappa_{\rm seismic,hth, ET}(f)|^2 &N_{{\rm seismic,h, ET}_1{\rm ET}_2}\\
        + |\Kappa_{\rm seismic,tth, ET}(f)|^2 &N_{{\rm seismic,t, ET}_1{\rm ET}_2},\\        
\mbox{where~~}      N_{{\rm seismic,v, ET}_1{\rm ET}_2} =& \frac{|{\rm CSD}_{\rm seismic,vertical}|}{\gamma_{{\rm ET}_1{\rm ET}_2}(f)S_0(f)}\\
\mbox{and~~}      N_{{\rm seismic,h, ET}_1{\rm ET}_2} =& \frac{|{\rm CSD}_{\rm seismic,horizontal}|}{\gamma_{{\rm ET}_1{\rm ET}_2}(f)S_0(f)}\\
\mbox{and~~}      N_{{\rm seismic,t, ET}_1{\rm ET}_2} =& \frac{|{\rm CSD}_{\rm seismic,tilt}|}{\gamma_{{\rm ET}_1{\rm ET}_2}(f)S_0(f)}\\
\mbox{and~~}   \\
        \hat{C}_{{\rm NN},{\rm ET}_1{\rm ET}_2}(f) =& N_{{\rm NN, ET}_1{\rm ET}_2}, \\
\mbox{where~~}      N_{{\rm NN, ET}_1{\rm ET}_2} =& \frac{{\rm S}_{\rm NN}}{\gamma_{{\rm ET}_1{\rm ET}_2}(f)S_0(f)}.\\
    \end{aligned}
\end{equation}
Here $\Kappa_{\rm seismic,vth, ET}(f)$, $\Kappa_{\rm seismic,hth, ET}(f)$ and $\Kappa_{\rm seismic,tth, ET}(f)$ are the three different seismic coupling functions. They describe the coupling of vertical seismic motions to a horizontal motion of the test mass (vth), horizontal seismic motion to a horizontal motion of the test mass (hth) and a tilt seismic motion to a horizontal motion of the test mass (tth), respectively. Please note that in this paper we will assume these three terms are uncorrelated. Whereas this is a good assumption, this assumption should be validated in future work.

In this paper we only have measurements of the correlated vertical and horizontal fields and therefore we neglect all effects coming from the tilt. Furthermore we assume identical seismic coupling for the three different interferometers ${\rm ET}_1, {\rm ET}_2, {\rm ET}_3$. In Eq. \ref{eq:UL_seismicTF} will provide the formalism to calculate upper limits on the seismic coupling function of which the results will be presented in Sec. \ref{sec:Results}. 

Since the calculated NN is the direct effect from the gravity fluctuations on the strain there is no additional coupling function ``$\Kappa_{\rm NN, ET}(f)$'' to take into account. The NN in Eq. \ref{eq:C_Seis_NN} can either come from Rayleigh waves or body waves, for which ${\rm S}_{\rm NN}$ was presented in Sec. \ref{sec:CorrNN}, respectively Eq. \ref{eq:Rayleigh} and Eq. \ref{eq:BWNN}.

The correlated seismic noise in the form of a cross spectral density are the quantities we presented in Sec. \ref{sec:CorrSeisNoise}. Here we use the absolute value of the CSD to be conservative. Furthermore, for the horizontal seismic CSD we calculate the ``omni-directional'' seismic CSD, where we take into account all possible cross-correlation combinations between the seismometer pairs, similar to the earlier study of magnetic fields \cite{PhysRevD.104.122006}, 
\begin{equation}
    \label{eq:OmnidirectionalCSD}
    \begin{aligned}
    {\rm CSD}_{\rm seismic,horizontal} =& \\
    \left[ \right.  |CSD_{{\rm seismic, NS-NS}}(f)|^2 +& |CSD_{{\rm seismic, NS-EW}}(f)|^2  \\
    + |CSD_{{\rm seismic, EW-NS}}(f)|^2 &+  |CSD_{{\rm seismic, EW-EW}}(f)|^2 \left.\right] ^{1/2}~,\\
    \end{aligned}
\end{equation}
where NS and EW indicate the orientation of the seismometers. Moreover, when calculating the upper limits for the seismic coupling function we will also introduce a factor of 5, where we assume incoherent sum of the possible coupling locations A-E as explained in Sec. \ref{sec:CorrNN} and Fig. \ref{fig:ETDiagram}.

The sensitivity of a search for an isotropic GWB can be related to the instantaneous sensitivity of the ET interferometer, referred to as the one-sided amplitude spectral density (ASD) $P_{\rm ET}(f)$, as follows~\cite{PhysRevD.46.5250,PhysRevD.59.102001,LivingRevRelativ20}:

\begin{equation}
    \label{eq:sigmaGWB}
    \sigma_{{\rm ET}_1{\rm ET}_2}(f) \approx \sqrt{\frac{1}{2T_{\rm obs}\Delta f}\frac{P_{\rm ET}^2(f)}{\gamma_{{\rm ET}_1{\rm ET}_2}^2(f)S_0^2(f)}}~,
\end{equation}
with $\Delta f$ the frequency resolution. Here we have assumed identical sensitivity in the different ET interferometers ${\rm ET}_1, {\rm ET}_2, {\rm ET}_3$. $\sigma_{{\rm ET}_1{\rm ET}_2}(f)$ is the standard deviation on the cross-correlation statistic defined in Eq.~\ref{eq:cross-correlationstatistic}, in the small signal-to-noise ratio (SNR) limit. Since the GWB one tries to observe is very weak this is a realistic assumption. 

Many of the expected signals for an isotropic GWB behave as a power-law. Therefore a more appropriate sensitivity to such a signal than $\sigma_{{\rm ET}_1{\rm ET}_2}(f)$ would be one that takes into account this broadband character of the expected signal. Such a broadband sensitivity is given by the so called power-law integrated (PI) curve: $\Omega^{\rm PI}_{{\rm ET}_1{\rm ET}_2}(f)$. $\Omega^{\rm PI}_{{\rm ET}_1{\rm ET}_2}(f)$ is constructed using $\sigma_{{\rm ET}_1{\rm ET}_2}(f)$ such that at any frequency a power-law signal $\Omega_{\rm GW}(f)$ with an SNR of 1 for the ${\rm ET}_1{\rm ET}_2$ baseline is tangent to this PI-sensitivity curve~\cite{Thrane:2013oya}. This makes $\Omega^{\rm PI}_{{\rm ET}_1{\rm ET}_2}(f)$ the relevant figure of merit to identify correlated broadband noise sources that could impact the search for an isotopic GWB.

Therefore we will equate $\Omega^{\rm PI}_{{\rm ET}_1{\rm ET}_2}(f)$ to the seismic cross-correlation statistic introduced in Eq. \ref{eq:C_Seis_NN} to compute the upper limits on the seismic coupling functions $\Kappa_{\rm seismic,vth, ET}(f)$ and $\Kappa_{\rm seismic,hth, ET}(f)$. We will construct these upper limits independently for vertical and horizontal seismic fields:

\begin{equation}
    \label{eq:UL_seismicTF}
    \begin{aligned}
        \Kappa_{\rm seismic,vth, ET}(f) \equiv& \sqrt{ \frac{\Omega^{\rm PI}_{{\rm ET}_1{\rm ET}_2}}{N_{{\rm seismic,v, ET}_1{\rm ET}_2}}} \\
        \Kappa_{\rm seismic,hth, ET}(f) \equiv& \sqrt{ \frac{\Omega^{\rm PI}_{{\rm ET}_1{\rm ET}_2}}{N_{{\rm seismic,h, ET}_1{\rm ET}_2}}}. 
    \end{aligned}
\end{equation}

\section{Noise projection - results}
\label{sec:Results}

Earlier work has studied the effect from seismic and Newtonian noise on the instantaneous sensitivity $P_{\rm ET}(f)$ \cite{Amann:2020jgo,Bader_2022,harms2022lower}. It was found that seismic noise could be dominant up to $\sim$ 2\,Hz-3\,Hz rather than just below 2\,Hz \cite{Amann:2020jgo}. To achieve the desired sensitivity, the effect of NN on the other hand should be reduced by about a factor of 3, which is considered to be feasible in case enough seismometers are deployed to apply effective (offline) noise mitigation \cite{Amann:2020jgo,Badaracco_2019}. 
In this paper we will only focus on the impact on the search for an isotropic GWB.

\begin{figure}
    \centering
    \includegraphics[width=\linewidth]{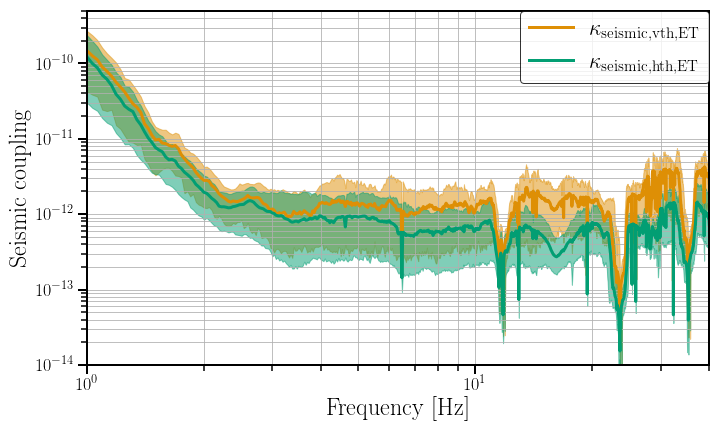}
    \caption{Upper limits on the seismic coupling functions $\protect\Kappa_{\protect\rm {seismic,vth, ET}}(f)$ and $\protect\Kappa_{\protect\rm {seismic,hth, ET}}(f)$ introduced in Sec. \ref{sec:NoiseProjection}. These upper limits are based on the 1-year power law integrated sensitivity $\Omega^{\protect\rm PI}_{\protect\rm ET}(f)$.}
    \label{fig:SeismicUL}
\end{figure}

Fig. \ref{fig:SeismicUL} represents the upper limits on the seismic coupling functions $\Kappa_{\rm seismic,vth, ET}(f)$ and $\Kappa_{\rm seismic,hth, ET}(f)$ such that seismic noise does not affect the broadband sensitivity $\Omega^{\rm PI}_{ET}(f)$ for the search for an isotropic GWB using one year of data. These upper limits were constructed using the seismic correlations measured underground at the former Homestake mine.
The design of the suspensions, and therefore the seismic coupling function, for the ET are still under investigation. The upper limits derived here can help making informed decisions on the necessary suspension requirements.

As a comparison, one can look at the 17\,m suspension design in \cite{ETdesignRep}, where the Fig 6.12 represents the seismic coupling for the hth component. Although this design will not be used in the final design of the ET, it gives an order of magnitude estimate of what can be expected. This indicates the seismic coupling might be too large up to a couple of Hz, but between 3Hz and 4Hz the seismic coupling function presented in \cite{ETdesignRep} reaches the desired level of $\sim$ $10^{-12}$.

Please note that by treating vertical and horizontal seismic correlated noise independently (and neglecting tilt) there is some room for residual contamination. For future designs of the ET suspensions dedicated follow-up studies should indicate whether they sufficiently reduce seismic (correlated) noise or not.

\begin{figure}
    \centering
    \includegraphics[width=\linewidth]{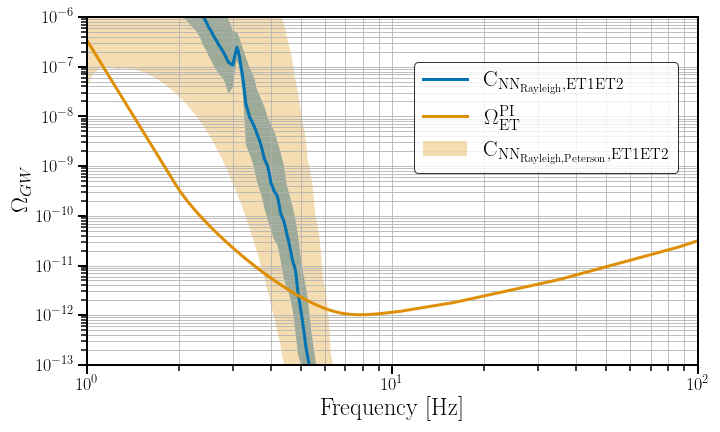}
    \caption{The projected impact from correlated NN from Rayleigh waves, as calculated in Sec. \ref{sec:CorrNN}. The blue line represents the median value and the associated surface is delimited by the 10\% and 90 $\%$ percentiles. We only show the projection for the NN which includes the attenuation gained by placing the ET 300m below the surface. As a comparison we also make the same projection using the Peterson low noise and high noise models. For the broadband ($\Omega^{\rm PI}_{{\rm ET}}$) sensitivity to a GWB we assumed one year of observation time (100\% duty cycle).}
    \label{fig:StochBudget_Terziet_NN}
\end{figure}

Fig. \ref{fig:StochBudget_Terziet_NN} projects the impact of NN from Rayleigh waves using the Terziet seismic data assuming a depth of 300\,m below the surface. Whereas the impact on the instantaneous sensitivity was shown to be problematic up to $\sim$ 3Hz in Fig. \ref{fig:NNTerziet400mAll}, the effect on the search for an isotropic GWB is affected up to $\sim$ 5\,Hz. At higher frequencies the depth of 300\,m ensures enough suppression of the NN from Rayleigh waves to prevent significant impact. For comparison also the NN from Rayleigh waves is shown using the Peterson low and high noise models.

We indicated earlier more dedicated (site-specific) studies are needed to take the local geology and anisotropies in the seismic field into account such that a more accurate noise projection can be achieved. However the Peterson high noise model predicts NN from Raleigh waves can contaminate the search for an isotropic GWB up to $\sim$ 6\,Hz, whereas this scenario represents the very worst case possibility: a very noise environment as well as 100\% correlated data. This indicates that regardless of the outcome of more precise site-specific studies investigating the correlations of seismic spectra at the surface, their impact through NN from Raleigh waves on the search for an isotropic GWB is expected to not exceed $\sim$ 6\,Hz, if the ET is located 300m below the surface.

\begin{figure}
    \centering
    \includegraphics[width=\linewidth]{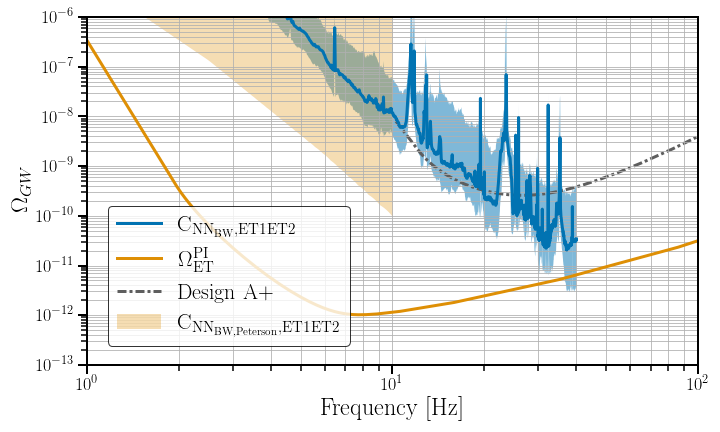}
    \caption{The projected impact from correlated NN from body-waves, as calculated in Sec. \ref{sec:CorrNN}. The blue line represents the median value and the associated surface is delimited by the 10\% and 90 $\%$ percentiles. As a comparison we also make the same projection using the Peterson low noise and high noise models. For the broadband ($\Omega^{\rm PI}_{{\rm ET}}$) sensitivity to a GWB we assumed one year of observation time (100\% duty cycle). The one year PI curve of the A+ design for the LIGO Hanford LIGO Livingston and Virgo detectors is represented by the dot-dashed curve. This curve was obtained using the open data provided by the LVK collaborations \cite{O3IsotropicDataset} and was first presented in \cite{KAGRA:2021kbb}. Please note: in this paper we present the 1$\sigma$ PI-curve, whereas in \cite{KAGRA:2021kbb} the 2$\sigma$ PI-curve is shown.}
    \label{fig:StochBudget_Homestake_NN}
\end{figure}

Fig. \ref{fig:StochBudget_Homestake_NN} on the other hand, displays a worrisome level of correlated NN from body waves which affects the search for an isotropic GWB up to at least 40\,Hz, the highest frequency with reliable data from the Homestake mine. Whereas the effect from Rayleigh waves decreases with the depth, there is no such reduction present for the NN from body waves. This leads to levels of correlated noise which are up to $\sim 8\times 10^6$ (90\% percentile), $\sim 6\times 10^5$ (50\% percentile) times larger than the desired sensitivity at $\sim$ 3\,Hz. Even when one would consider the seismic correlated noise observed during the night at Homestake, this does not alter significantly the impact displayed in Fig. \ref{fig:StochBudget_Homestake_NN}. As mentioned in Sec. \ref{sec:CorrSeisNoise} the 90\% percentile observed during the night is quite similar to the 50\% percentile observed during the day and the correlated NN from body-waves still affects the isotropic search for a SGWB up to 40Hz.

Here we would like to point out that the local seismic environment of the ET candidate sites might give different results than the ones obtained using the data from the Homestake mine. To this extent site-specific studies of underground correlation measurements over the scale of several hundreds of meters could give more insights. Evidently in case the ET candidate site would have lower levels of correlated seismic noise than observed at Homestake, this will lead to less stringent constraints. However we also want to note that infrastructure for the ET installed in the underground environment will create local seismic fields that could be louder and/or more strongly correlated at the mirrors of the different interferometers. To this extent it is also crucial to look into the possible disturbances of infrastructure and methods to reduce their effect. Also should be understood whether the effect from infrastructure or ambient seismic environment interplay and which is the dominant factor.

Noise subtraction for NN is being investigated \cite{PhysRevD.86.102001,PhysRevD.92.022001,Coughlin_2014_seismic,Coughlin_2016,PhysRevLett.121.221104,Tringali_2019,Badaracco_2019,Badaracco_2020,NN_Sardinia2020,10.1785/0220200186,Bader_2022,Koley_2022}, however with a focus to reduce the impact needed to reach the desired ASD sensitivity. The amount of reduction needed for the search for an isotropic GWB is many orders of magnitude higher. However, the budget shown in Fig. \ref{fig:StochBudget_Homestake_NN} is at the level of the baseline, such that the improvement of 5-7 orders of magnitude at $\sim$ 3\;Hz corresponds to an improvement of about 3 orders of magnitude at every interferometer. Reducing the impact of NN on the ET with such levels is nevertheless considered to be out of reach. A more realistic (and already optimistic) level of reduction is a factor of 10 at each interferometer \cite{Badaracco_2019}, that is a factor of 100 for the baseline. Even with this amount of NN subtraction, NN from body waves could be expected to contaminate the search for an isotropic GWB up to $\sim$ 30Hz. This would not only imply we lose the most sensitive region of the analysis but also that in the low frequency region no to negligible improvement is gained compared to the expected sensitivity reachable by the LIGO and Virgo instruments after their A+ upgrades \cite{APlussReference}. This will have a dramatic impact on the search for an isotropic GWB, regardless of the excepted source. As an example for the GWB coming from unresolved CBC events, more than 95\% of the SNR is expected to be below 30\,Hz for the Xylophone ET design considered here \cite{PhysRevD.89.084046}.

\section{Conclusion}
\label{sec:Conclusion}

The ET promises to be a powerful instrument to observe GWs in the coming decades, outperforming LIGO and Virgo. The ET is also planned to have an unprecedented sensitivity to GWs in the 'low-frequency' region of several Hz to several tens of Hz \cite{Maggiore:2019uih}. This could significantly improve the capability of the ET to observe a GWB of either astrophysical or cosmological origin. To illustrate, more than 95\% of ETs sensitivity to a GWB coming from unresolved CBC events is from below 30Hz \cite{PhysRevD.89.084046}.

However this low frequency region is also susceptible to several correlated noise sources.
An earlier study showed correlated magnetic fields could affect GWB searches up to $\sim$ 30\,Hz and magnetic coupling functions should be reduced by several orders of magnitude compared to the observed magnetic coupling at LIGO and Virgo \cite{PhysRevD.104.122006}.
In the same context we have investigated the possible impact from correlated seismic and Newtonian noise on the ET and its search for a GWB.

We analysed correlations between vertical seismic spectra measured between sensors of the Terziet surface array \cite{3T_network}, with a horizontal separation of several hundreds of meters. Furthermore, both vertical and horizontal seismic spectra were analysed for data taken $\sim$ 610\,m below the surface at the former Homestake mine \cite{10.1785/0220170228}.
At both locations significant seismic coherence was observed up to 10\,Hz, 50\% of the time or more. Also between 10\,Hz and 50\,Hz significant coherence was observed. Although the observed correlated seismic spectra above $\sim$ 10\,Hz are lower during the nights due to less human activity, no decrease in seismic coherence is observed during the weekend.

By comparing correlated seismic spectra observed at different horizontal sensor separation, ranging from $\sim$ 200\,m to $\sim$ 810\,m, we have tried to identify the possible impact from the distance between the test masses from two different ET interferometers. Currentely, this distance is foreseen to be about 300\,m to 500\,m. The observed correlated seismic spectra using data from the Terziet surface array do vary significantly above 1 \,Hz. However, no pattern related with the distance between sensors is observed, leading to the conclusion that the variation is most likely due to local variations in the number and location of the noise sources compared to the sensor locations. Site-specific studies using the entire array of deployed sensors should give further insights. However, the data presented here is able to give us an order of magnitude of the correlated noise and its impact on the ET. The observed seismic spectra from the more quiet Homestake mine, show no dependence on horizontal separation between the sensors. The correlated seismic spectra at Homestake in the vertical direction are also very similar to the horizontal seismic spectra if one looks at aligned sensors. The correlated seismic spectra between two perpendicular seismometers is lower up to $\sim$ 2-3\,Hz when comparing to two aligned seismometers. At higher frequencies no difference is observed.

The measurements at the surface (Terziet) are used to predict the levels of correlated Newtonian noise from Rayleigh waves. If the ET were built at the surface, the NN would have been roughly one order of magnitude above ETs design sensitivity in the region of 2\,Hz - 10\,Hz. Please note that in these calculations we used a simplified model not taking into account the entire geological complexity, as was done in \cite{Bader_2022}. By building the ET 300\,m underground a marginal effect on the ETs ASD is present up to 3\,Hz.
Similarly we used the underground measurements (Homestake) to calculate the level of correlated NN from body waves, of which the median is 3-5 times larger than the ETs design sensitivity between 2\,Hz and  10\,Hz.

The key result of this paper is the propagation of these correlated seismic and Newtonian noise estimates onto the sensitivity for the search for an isotropic GWB. First of all, we presented upper limits on the seismic coupling function which has to be of the order of $\sim 10^{-12}$ or lower above 2\,Hz. Secondly, the effect from correlated NN from Rayleigh waves is found to impact the GWB sensitivity up to $\sim$ 5\,Hz, in the scenario the ET facility is located 300\,m below the surface. Finally, correlated NN from body waves are found to affect the GWB sensitivity up to 40\,Hz, with a maximal effect of $\sim 8\times 10^6$ (90\% percentile), $\sim 6\times 10^5$ (50\% percentile) at 3\,Hz. Even with a NN mitigation factor of 10 in each interferometer, the GWB search would be contaminated by correlated noise from body waves below 30\,Hz.
Even when considering the seismic correlations during the local night, when minimal anthropogenic activities take place the impact on the stochastic search remains of the same order of magnitude, i.e. $\sim 10^5-10^6$ at 3Hz and a non-negligible impact up to 40Hz.

The presented results for the effect of correlated NN indicate a negative impact on the scientific goals for the ET. This subject must be further studied, and subsequent work has to include simulations for a more accurate prediction of the NN. 
These studies could also be envisioned to include site-specific studies to get a more accurate understanding of the effect at the candidate sites. However, not only the correlated ambient seismic fields at the sites should be understood, since the ET infrastructure could lead to additional local (and possibly correlated) seismic fields. One should try to understand whether this effect is dominant over the sites ambient seismic correlations and if so to which extent clever design and placement of the infrastructure could decrease this impact.
Other studies should try to understand the interplay between the correlated NN and the ET interferometric system. An accurate simulation of seismic fields interacting on the ET system should be performed to understand the possible interplay of the multple coupling locations of seismic and Newtonian noise as introduced in Fig. \ref{fig:ETDiagram}
Finally, one can also investigate the possibility of using the null channel as a tool to estimate correlated noise. A recent paper \cite{2022arXiv220500416J} proposes a formalism to estimate the ETs noise PSD which also gives insight in the cross-correlated noise terms. However, more work has to be done to understand its efficiency in scenario's with realistic correlated noise sources, such as the seismic and Newtonian noise as presented here. 

\acknowledgements

We thank Jan Harms for constructive comments. The authors acknowledge access to computational resources provided by the LIGO Laboratory supported by National Science Foundation Grants PHY-0757058 and PHY-0823459. GB thanks the laboratory  Artemis, Observatoire de la C\^ote d'Azur, for hospitality and welcome.

This paper has been given LIGO DCC number P2200172, Virgo TDS number VIR-0589A-22 and ET TDS number 	ET-0124A-22.

K.J. is supported by FWO-Vlaanderen via grant number 11C5720N.

\bibliography{references}

\begin{thebibliography}{60}%
\makeatletter
\providecommand \@ifxundefined [1]{%
 \@ifx{#1\undefined}
}%
\providecommand \@ifnum [1]{%
 \ifnum #1\expandafter \@firstoftwo
 \else \expandafter \@secondoftwo
 \fi
}%
\providecommand \@ifx [1]{%
 \ifx #1\expandafter \@firstoftwo
 \else \expandafter \@secondoftwo
 \fi
}%
\providecommand \natexlab [1]{#1}%
\providecommand \enquote  [1]{``#1''}%
\providecommand \bibnamefont  [1]{#1}%
\providecommand \bibfnamefont [1]{#1}%
\providecommand \citenamefont [1]{#1}%
\providecommand \href@noop [0]{\@secondoftwo}%
\providecommand \href [0]{\begingroup \@sanitize@url \@href}%
\providecommand \@href[1]{\@@startlink{#1}\@@href}%
\providecommand \@@href[1]{\endgroup#1\@@endlink}%
\providecommand \@sanitize@url [0]{\catcode `\\12\catcode `\$12\catcode
  `\&12\catcode `\#12\catcode `\^12\catcode `\_12\catcode `\%12\relax}%
\providecommand \@@startlink[1]{}%
\providecommand \@@endlink[0]{}%
\providecommand \url  [0]{\begingroup\@sanitize@url \@url }%
\providecommand \@url [1]{\endgroup\@href {#1}{\urlprefix }}%
\providecommand \urlprefix  [0]{URL }%
\providecommand \Eprint [0]{\href }%
\providecommand \doibase [0]{http://dx.doi.org/}%
\providecommand \selectlanguage [0]{\@gobble}%
\providecommand \bibinfo  [0]{\@secondoftwo}%
\providecommand \bibfield  [0]{\@secondoftwo}%
\providecommand \translation [1]{[#1]}%
\providecommand \BibitemOpen [0]{}%
\providecommand \bibitemStop [0]{}%
\providecommand \bibitemNoStop [0]{.\EOS\space}%
\providecommand \EOS [0]{\spacefactor3000\relax}%
\providecommand \BibitemShut  [1]{\csname bibitem#1\endcsname}%
\let\auto@bib@innerbib\@empty
\bibitem [{\citenamefont {Christensen}(2018)}]{Christensen_2018}%
  \BibitemOpen
  \bibfield  {author} {\bibinfo {author} {\bibfnamefont {N.}~\bibnamefont
  {Christensen}},\ }\href {\doibase 10.1088/1361-6633/aae6b5} {\bibfield
  {journal} {\bibinfo  {journal} {Reports on Progress in Physics}\ }\textbf
  {\bibinfo {volume} {82}},\ \bibinfo {pages} {016903} (\bibinfo {year}
  {2018})}\BibitemShut {NoStop}%
\bibitem [{\citenamefont {Aasi}\ \emph
  {et~al.}(2015{\natexlab{a}})\citenamefont {Aasi} \emph {et~al.}}]{2015}%
  \BibitemOpen
  \bibfield  {author} {\bibinfo {author} {\bibfnamefont {J.}~\bibnamefont
  {Aasi}} \emph {et~al.},\ }\href {\doibase 10.1088/0264-9381/32/7/074001}
  {\bibfield  {journal} {\bibinfo  {journal} {Classical and Quantum Gravity}\
  }\textbf {\bibinfo {volume} {32}},\ \bibinfo {pages} {074001} (\bibinfo
  {year} {2015}{\natexlab{a}})}\BibitemShut {NoStop}%
\bibitem [{\citenamefont {Acernese}\ \emph {et~al.}(2015)\citenamefont
  {Acernese} \emph {et~al.}}]{VIRGO:2014yos}%
  \BibitemOpen
  \bibfield  {author} {\bibinfo {author} {\bibfnamefont {F.}~\bibnamefont
  {Acernese}} \emph {et~al.} (\bibinfo {collaboration} {VIRGO}),\ }\href
  {\doibase 10.1088/0264-9381/32/2/024001} {\bibfield  {journal} {\bibinfo
  {journal} {Class. Quant. Grav.}\ }\textbf {\bibinfo {volume} {32}},\ \bibinfo
  {pages} {024001} (\bibinfo {year} {2015})},\ \Eprint
  {http://arxiv.org/abs/1408.3978} {arXiv:1408.3978 [gr-qc]} \BibitemShut
  {NoStop}%
\bibitem [{\citenamefont {Aso}\ \emph {et~al.}(2013)\citenamefont {Aso},
  \citenamefont {Michimura}, \citenamefont {Somiya}, \citenamefont {Ando},
  \citenamefont {Miyakawa}, \citenamefont {Sekiguchi}, \citenamefont
  {Tatsumi},\ and\ \citenamefont {Yamamoto}}]{PhysRevD.88.043007}%
  \BibitemOpen
  \bibfield  {author} {\bibinfo {author} {\bibfnamefont {Y.}~\bibnamefont
  {Aso}}, \bibinfo {author} {\bibfnamefont {Y.}~\bibnamefont {Michimura}},
  \bibinfo {author} {\bibfnamefont {K.}~\bibnamefont {Somiya}}, \bibinfo
  {author} {\bibfnamefont {M.}~\bibnamefont {Ando}}, \bibinfo {author}
  {\bibfnamefont {O.}~\bibnamefont {Miyakawa}}, \bibinfo {author}
  {\bibfnamefont {T.}~\bibnamefont {Sekiguchi}}, \bibinfo {author}
  {\bibfnamefont {D.}~\bibnamefont {Tatsumi}}, \ and\ \bibinfo {author}
  {\bibfnamefont {H.}~\bibnamefont {Yamamoto}} (\bibinfo {collaboration} {The
  KAGRA Collaboration}),\ }\href {\doibase 10.1103/PhysRevD.88.043007}
  {\bibfield  {journal} {\bibinfo  {journal} {Phys. Rev. D}\ }\textbf {\bibinfo
  {volume} {88}},\ \bibinfo {pages} {043007} (\bibinfo {year}
  {2013})}\BibitemShut {NoStop}%
\bibitem [{\citenamefont {{Schumann}}(1952{\natexlab{a}})}]{Schumann1}%
  \BibitemOpen
  \bibfield  {author} {\bibinfo {author} {\bibfnamefont {W.~O.}\ \bibnamefont
  {{Schumann}}},\ }\href {\doibase 10.1515/zna-1952-0202} {\bibfield  {journal}
  {\bibinfo  {journal} {Zeitschrift Naturforschung Teil A}\ }\textbf {\bibinfo
  {volume} {7}},\ \bibinfo {pages} {149} (\bibinfo {year}
  {1952}{\natexlab{a}})}\BibitemShut {NoStop}%
\bibitem [{\citenamefont {{Schumann}}(1952{\natexlab{b}})}]{Schumann2}%
  \BibitemOpen
  \bibfield  {author} {\bibinfo {author} {\bibfnamefont {W.~O.}\ \bibnamefont
  {{Schumann}}},\ }\href {\doibase 10.1515/zna-1952-3-404} {\bibfield
  {journal} {\bibinfo  {journal} {Zeitschrift Naturforschung Teil A}\ }\textbf
  {\bibinfo {volume} {7}},\ \bibinfo {pages} {250} (\bibinfo {year}
  {1952}{\natexlab{b}})}\BibitemShut {NoStop}%
\bibitem [{\citenamefont {Thrane}\ \emph {et~al.}(2013)\citenamefont {Thrane},
  \citenamefont {Christensen},\ and\ \citenamefont
  {Schofield}}]{Thrane:2013npa}%
  \BibitemOpen
  \bibfield  {author} {\bibinfo {author} {\bibfnamefont {E.}~\bibnamefont
  {Thrane}}, \bibinfo {author} {\bibfnamefont {N.}~\bibnamefont {Christensen}},
  \ and\ \bibinfo {author} {\bibfnamefont {R.~M.~S.}\ \bibnamefont
  {Schofield}},\ }\href {\doibase 10.1103/PhysRevD.87.123009} {\bibfield
  {journal} {\bibinfo  {journal} {Phys. Rev. D}\ }\textbf {\bibinfo {volume}
  {87}},\ \bibinfo {pages} {123009} (\bibinfo {year} {2013})},\ \Eprint
  {http://arxiv.org/abs/1303.2613} {arXiv:1303.2613 [astro-ph.IM]} \BibitemShut
  {NoStop}%
\bibitem [{\citenamefont {Thrane}\ \emph {et~al.}(2014)\citenamefont {Thrane},
  \citenamefont {Christensen}, \citenamefont {Schofield},\ and\ \citenamefont
  {Effler}}]{Thrane:2014yza}%
  \BibitemOpen
  \bibfield  {author} {\bibinfo {author} {\bibfnamefont {E.}~\bibnamefont
  {Thrane}}, \bibinfo {author} {\bibfnamefont {N.}~\bibnamefont {Christensen}},
  \bibinfo {author} {\bibfnamefont {R.~M.~S.}\ \bibnamefont {Schofield}}, \
  and\ \bibinfo {author} {\bibfnamefont {A.}~\bibnamefont {Effler}},\ }\href
  {\doibase 10.1103/PhysRevD.90.023013} {\bibfield  {journal} {\bibinfo
  {journal} {Phys. Rev. D}\ }\textbf {\bibinfo {volume} {90}},\ \bibinfo
  {pages} {023013} (\bibinfo {year} {2014})},\ \Eprint
  {http://arxiv.org/abs/1406.2367} {arXiv:1406.2367 [astro-ph.IM]} \BibitemShut
  {NoStop}%
\bibitem [{\citenamefont {Coughlin}\ \emph
  {et~al.}(2016{\natexlab{a}})\citenamefont {Coughlin} \emph
  {et~al.}}]{Coughlin:2016vor}%
  \BibitemOpen
  \bibfield  {author} {\bibinfo {author} {\bibfnamefont {M.~W.}\ \bibnamefont
  {Coughlin}} \emph {et~al.},\ }\href {\doibase 10.1088/0264-9381/33/22/224003}
  {\bibfield  {journal} {\bibinfo  {journal} {Class. Quant. Grav.}\ }\textbf
  {\bibinfo {volume} {33}},\ \bibinfo {pages} {224003} (\bibinfo {year}
  {2016}{\natexlab{a}})},\ \Eprint {http://arxiv.org/abs/1606.01011}
  {arXiv:1606.01011 [gr-qc]} \BibitemShut {NoStop}%
\bibitem [{\citenamefont {Himemoto}\ and\ \citenamefont
  {Taruya}(2017)}]{Himemoto:2017gnw}%
  \BibitemOpen
  \bibfield  {author} {\bibinfo {author} {\bibfnamefont {Y.}~\bibnamefont
  {Himemoto}}\ and\ \bibinfo {author} {\bibfnamefont {A.}~\bibnamefont
  {Taruya}},\ }\href {\doibase 10.1103/PhysRevD.96.022004} {\bibfield
  {journal} {\bibinfo  {journal} {Phys. Rev. D}\ }\textbf {\bibinfo {volume}
  {96}},\ \bibinfo {pages} {022004} (\bibinfo {year} {2017})},\ \Eprint
  {http://arxiv.org/abs/1704.07084} {arXiv:1704.07084 [astro-ph.IM]}
  \BibitemShut {NoStop}%
\bibitem [{\citenamefont {Coughlin}\ \emph
  {et~al.}(2018{\natexlab{a}})\citenamefont {Coughlin}, \citenamefont {Cirone},
  \citenamefont {Meyers}, \citenamefont {Atsuta}, \citenamefont {Boschi},
  \citenamefont {Chincarini}, \citenamefont {Christensen}, \citenamefont
  {De~Rosa}, \citenamefont {Effler}, \citenamefont {Fiori}, \citenamefont
  {Go\l{}kowski}, \citenamefont {Guidry}, \citenamefont {Harms}, \citenamefont
  {Hayama}, \citenamefont {Kataoka}, \citenamefont {Kubisz}, \citenamefont
  {Kulak}, \citenamefont {Laxen}, \citenamefont {Matas}, \citenamefont
  {Mlynarczyk}, \citenamefont {Ogawa}, \citenamefont {Paoletti}, \citenamefont
  {Salvador}, \citenamefont {Schofield}, \citenamefont {Somiya},\ and\
  \citenamefont {Thrane}}]{Coughlin:2018tjc}%
  \BibitemOpen
  \bibfield  {author} {\bibinfo {author} {\bibfnamefont {M.~W.}\ \bibnamefont
  {Coughlin}}, \bibinfo {author} {\bibfnamefont {A.}~\bibnamefont {Cirone}},
  \bibinfo {author} {\bibfnamefont {P.}~\bibnamefont {Meyers}}, \bibinfo
  {author} {\bibfnamefont {S.}~\bibnamefont {Atsuta}}, \bibinfo {author}
  {\bibfnamefont {V.}~\bibnamefont {Boschi}}, \bibinfo {author} {\bibfnamefont
  {A.}~\bibnamefont {Chincarini}}, \bibinfo {author} {\bibfnamefont {N.~L.}\
  \bibnamefont {Christensen}}, \bibinfo {author} {\bibfnamefont
  {R.}~\bibnamefont {De~Rosa}}, \bibinfo {author} {\bibfnamefont
  {A.}~\bibnamefont {Effler}}, \bibinfo {author} {\bibfnamefont
  {I.}~\bibnamefont {Fiori}}, \bibinfo {author} {\bibfnamefont
  {M.}~\bibnamefont {Go\l{}kowski}}, \bibinfo {author} {\bibfnamefont
  {M.}~\bibnamefont {Guidry}}, \bibinfo {author} {\bibfnamefont
  {J.}~\bibnamefont {Harms}}, \bibinfo {author} {\bibfnamefont
  {K.}~\bibnamefont {Hayama}}, \bibinfo {author} {\bibfnamefont
  {Y.}~\bibnamefont {Kataoka}}, \bibinfo {author} {\bibfnamefont
  {J.}~\bibnamefont {Kubisz}}, \bibinfo {author} {\bibfnamefont
  {A.}~\bibnamefont {Kulak}}, \bibinfo {author} {\bibfnamefont
  {M.}~\bibnamefont {Laxen}}, \bibinfo {author} {\bibfnamefont
  {A.}~\bibnamefont {Matas}}, \bibinfo {author} {\bibfnamefont
  {J.}~\bibnamefont {Mlynarczyk}}, \bibinfo {author} {\bibfnamefont
  {T.}~\bibnamefont {Ogawa}}, \bibinfo {author} {\bibfnamefont
  {F.}~\bibnamefont {Paoletti}}, \bibinfo {author} {\bibfnamefont
  {J.}~\bibnamefont {Salvador}}, \bibinfo {author} {\bibfnamefont
  {R.}~\bibnamefont {Schofield}}, \bibinfo {author} {\bibfnamefont
  {K.}~\bibnamefont {Somiya}}, \ and\ \bibinfo {author} {\bibfnamefont
  {E.}~\bibnamefont {Thrane}},\ }\href {\doibase 10.1103/PhysRevD.97.102007}
  {\bibfield  {journal} {\bibinfo  {journal} {Phys. Rev. D}\ }\textbf {\bibinfo
  {volume} {97}},\ \bibinfo {pages} {102007} (\bibinfo {year}
  {2018}{\natexlab{a}})}\BibitemShut {NoStop}%
\bibitem [{\citenamefont {Himemoto}\ and\ \citenamefont
  {Taruya}(2019)}]{Himemoto:2019iwd}%
  \BibitemOpen
  \bibfield  {author} {\bibinfo {author} {\bibfnamefont {Y.}~\bibnamefont
  {Himemoto}}\ and\ \bibinfo {author} {\bibfnamefont {A.}~\bibnamefont
  {Taruya}},\ }\href {\doibase 10.1103/PhysRevD.100.082001} {\bibfield
  {journal} {\bibinfo  {journal} {Phys. Rev. D}\ }\textbf {\bibinfo {volume}
  {100}},\ \bibinfo {pages} {082001} (\bibinfo {year} {2019})},\ \Eprint
  {http://arxiv.org/abs/1908.10635} {arXiv:1908.10635 [astro-ph.IM]}
  \BibitemShut {NoStop}%
\bibitem [{\citenamefont {Meyers}\ \emph {et~al.}(2020)\citenamefont {Meyers},
  \citenamefont {Martinovic}, \citenamefont {Christensen},\ and\ \citenamefont
  {Sakellariadou}}]{Meyers:2020qrb}%
  \BibitemOpen
  \bibfield  {author} {\bibinfo {author} {\bibfnamefont {P.~M.}\ \bibnamefont
  {Meyers}}, \bibinfo {author} {\bibfnamefont {K.}~\bibnamefont {Martinovic}},
  \bibinfo {author} {\bibfnamefont {N.}~\bibnamefont {Christensen}}, \ and\
  \bibinfo {author} {\bibfnamefont {M.}~\bibnamefont {Sakellariadou}},\ }\href
  {\doibase 10.1103/PhysRevD.102.102005} {\bibfield  {journal} {\bibinfo
  {journal} {Phys. Rev. D}\ }\textbf {\bibinfo {volume} {102}},\ \bibinfo
  {pages} {102005} (\bibinfo {year} {2020})},\ \Eprint
  {http://arxiv.org/abs/2008.00789} {arXiv:2008.00789 [gr-qc]} \BibitemShut
  {NoStop}%
\bibitem [{\citenamefont {Janssens}\ \emph {et~al.}(2021)\citenamefont
  {Janssens}, \citenamefont {Martinovic}, \citenamefont {Christensen},
  \citenamefont {Meyers},\ and\ \citenamefont
  {Sakellariadou}}]{PhysRevD.104.122006}%
  \BibitemOpen
  \bibfield  {author} {\bibinfo {author} {\bibfnamefont {K.}~\bibnamefont
  {Janssens}}, \bibinfo {author} {\bibfnamefont {K.}~\bibnamefont
  {Martinovic}}, \bibinfo {author} {\bibfnamefont {N.}~\bibnamefont
  {Christensen}}, \bibinfo {author} {\bibfnamefont {P.~M.}\ \bibnamefont
  {Meyers}}, \ and\ \bibinfo {author} {\bibfnamefont {M.}~\bibnamefont
  {Sakellariadou}},\ }\href {\doibase 10.1103/PhysRevD.104.122006} {\bibfield
  {journal} {\bibinfo  {journal} {Phys. Rev. D}\ }\textbf {\bibinfo {volume}
  {104}},\ \bibinfo {pages} {122006} (\bibinfo {year} {2021})}\BibitemShut
  {NoStop}%
\bibitem [{\citenamefont {Aasi}\ \emph
  {et~al.}(2015{\natexlab{b}})\citenamefont {Aasi} \emph
  {et~al.}}]{LIGOScientific:2014sej}%
  \BibitemOpen
  \bibfield  {author} {\bibinfo {author} {\bibfnamefont {J.}~\bibnamefont
  {Aasi}} \emph {et~al.} (\bibinfo {collaboration} {LIGO Scientific, VIRGO}),\
  }\href {\doibase 10.1103/PhysRevD.91.022003} {\bibfield  {journal} {\bibinfo
  {journal} {Phys. Rev. D}\ }\textbf {\bibinfo {volume} {91}},\ \bibinfo
  {pages} {022003} (\bibinfo {year} {2015}{\natexlab{b}})},\ \Eprint
  {http://arxiv.org/abs/1410.6211} {arXiv:1410.6211 [gr-qc]} \BibitemShut
  {NoStop}%
\bibitem [{\citenamefont {Punturo}\ \emph {et~al.}(2010)\citenamefont {Punturo}
  \emph {et~al.}}]{Punturo:2010zz}%
  \BibitemOpen
  \bibfield  {author} {\bibinfo {author} {\bibfnamefont {M.}~\bibnamefont
  {Punturo}} \emph {et~al.},\ }\href {\doibase 10.1088/0264-9381/27/19/194002}
  {\bibfield  {journal} {\bibinfo  {journal} {Class. Quant. Grav.}\ }\textbf
  {\bibinfo {volume} {27}},\ \bibinfo {pages} {194002} (\bibinfo {year}
  {2010})}\BibitemShut {NoStop}%
\bibitem [{\citenamefont {Hild}\ \emph {et~al.}(2011)\citenamefont {Hild} \emph
  {et~al.}}]{Hild:2010id}%
  \BibitemOpen
  \bibfield  {author} {\bibinfo {author} {\bibfnamefont {S.}~\bibnamefont
  {Hild}} \emph {et~al.},\ }\href {\doibase 10.1088/0264-9381/28/9/094013}
  {\bibfield  {journal} {\bibinfo  {journal} {Class. Quant. Grav.}\ }\textbf
  {\bibinfo {volume} {28}},\ \bibinfo {pages} {094013} (\bibinfo {year}
  {2011})},\ \Eprint {http://arxiv.org/abs/1012.0908} {arXiv:1012.0908 [gr-qc]}
  \BibitemShut {NoStop}%
\bibitem [{\citenamefont {Amann}\ \emph {et~al.}(2020)\citenamefont {Amann}
  \emph {et~al.}}]{Amann:2020jgo}%
  \BibitemOpen
  \bibfield  {author} {\bibinfo {author} {\bibfnamefont {F.}~\bibnamefont
  {Amann}} \emph {et~al.},\ }\href {\doibase 10.1063/5.0018414} {\bibfield
  {journal} {\bibinfo  {journal} {Rev. Sci. Instrum.}\ }\textbf {\bibinfo
  {volume} {91}},\ \bibinfo {pages} {9} (\bibinfo {year} {2020})},\ \Eprint
  {http://arxiv.org/abs/2003.03434} {arXiv:2003.03434 [physics.ins-det]}
  \BibitemShut {NoStop}%
\bibitem [{\citenamefont {Driggers}\ \emph
  {et~al.}(2012{\natexlab{a}})\citenamefont {Driggers}, \citenamefont {Harms},\
  and\ \citenamefont {Adhikari}}]{PhysRevD.86.102001}%
  \BibitemOpen
  \bibfield  {author} {\bibinfo {author} {\bibfnamefont {J.~C.}\ \bibnamefont
  {Driggers}}, \bibinfo {author} {\bibfnamefont {J.}~\bibnamefont {Harms}}, \
  and\ \bibinfo {author} {\bibfnamefont {R.~X.}\ \bibnamefont {Adhikari}},\
  }\href {\doibase 10.1103/PhysRevD.86.102001} {\bibfield  {journal} {\bibinfo
  {journal} {Phys. Rev. D}\ }\textbf {\bibinfo {volume} {86}},\ \bibinfo
  {pages} {102001} (\bibinfo {year} {2012}{\natexlab{a}})}\BibitemShut
  {NoStop}%
\bibitem [{\citenamefont {Harms}\ and\ \citenamefont
  {Paik}(2015)}]{PhysRevD.92.022001}%
  \BibitemOpen
  \bibfield  {author} {\bibinfo {author} {\bibfnamefont {J.}~\bibnamefont
  {Harms}}\ and\ \bibinfo {author} {\bibfnamefont {H.~J.}\ \bibnamefont
  {Paik}},\ }\href {\doibase 10.1103/PhysRevD.92.022001} {\bibfield  {journal}
  {\bibinfo  {journal} {Phys. Rev. D}\ }\textbf {\bibinfo {volume} {92}},\
  \bibinfo {pages} {022001} (\bibinfo {year} {2015})}\BibitemShut {NoStop}%
\bibitem [{\citenamefont {Coughlin}\ \emph {et~al.}(2014)\citenamefont
  {Coughlin}, \citenamefont {Harms}, \citenamefont {Christensen}, \citenamefont
  {Dergachev}, \citenamefont {DeSalvo}, \citenamefont {Kandhasamy},\ and\
  \citenamefont {Mandic}}]{Coughlin_2014_seismic}%
  \BibitemOpen
  \bibfield  {author} {\bibinfo {author} {\bibfnamefont {M.}~\bibnamefont
  {Coughlin}}, \bibinfo {author} {\bibfnamefont {J.}~\bibnamefont {Harms}},
  \bibinfo {author} {\bibfnamefont {N.}~\bibnamefont {Christensen}}, \bibinfo
  {author} {\bibfnamefont {V.}~\bibnamefont {Dergachev}}, \bibinfo {author}
  {\bibfnamefont {R.}~\bibnamefont {DeSalvo}}, \bibinfo {author} {\bibfnamefont
  {S.}~\bibnamefont {Kandhasamy}}, \ and\ \bibinfo {author} {\bibfnamefont
  {V.}~\bibnamefont {Mandic}},\ }\href {\doibase
  10.1088/0264-9381/31/21/215003} {\bibfield  {journal} {\bibinfo  {journal}
  {Classical and Quantum Gravity}\ }\textbf {\bibinfo {volume} {31}},\ \bibinfo
  {pages} {215003} (\bibinfo {year} {2014})}\BibitemShut {NoStop}%
\bibitem [{\citenamefont {Coughlin}\ \emph
  {et~al.}(2016{\natexlab{b}})\citenamefont {Coughlin}, \citenamefont {Mukund},
  \citenamefont {Harms}, \citenamefont {Driggers}, \citenamefont {Adhikari},\
  and\ \citenamefont {Mitra}}]{Coughlin_2016}%
  \BibitemOpen
  \bibfield  {author} {\bibinfo {author} {\bibfnamefont {M.}~\bibnamefont
  {Coughlin}}, \bibinfo {author} {\bibfnamefont {N.}~\bibnamefont {Mukund}},
  \bibinfo {author} {\bibfnamefont {J.}~\bibnamefont {Harms}}, \bibinfo
  {author} {\bibfnamefont {J.}~\bibnamefont {Driggers}}, \bibinfo {author}
  {\bibfnamefont {R.}~\bibnamefont {Adhikari}}, \ and\ \bibinfo {author}
  {\bibfnamefont {S.}~\bibnamefont {Mitra}},\ }\href {\doibase
  10.1088/0264-9381/33/24/244001} {\bibfield  {journal} {\bibinfo  {journal}
  {Classical and Quantum Gravity}\ }\textbf {\bibinfo {volume} {33}},\ \bibinfo
  {pages} {244001} (\bibinfo {year} {2016}{\natexlab{b}})}\BibitemShut
  {NoStop}%
\bibitem [{\citenamefont {Coughlin}\ \emph
  {et~al.}(2018{\natexlab{b}})\citenamefont {Coughlin}, \citenamefont {Harms},
  \citenamefont {Driggers}, \citenamefont {McManus}, \citenamefont {Mukund},
  \citenamefont {Ross}, \citenamefont {Slagmolen},\ and\ \citenamefont
  {Venkateswara}}]{PhysRevLett.121.221104}%
  \BibitemOpen
  \bibfield  {author} {\bibinfo {author} {\bibfnamefont {M.~W.}\ \bibnamefont
  {Coughlin}}, \bibinfo {author} {\bibfnamefont {J.}~\bibnamefont {Harms}},
  \bibinfo {author} {\bibfnamefont {J.}~\bibnamefont {Driggers}}, \bibinfo
  {author} {\bibfnamefont {D.~J.}\ \bibnamefont {McManus}}, \bibinfo {author}
  {\bibfnamefont {N.}~\bibnamefont {Mukund}}, \bibinfo {author} {\bibfnamefont
  {M.~P.}\ \bibnamefont {Ross}}, \bibinfo {author} {\bibfnamefont {B.~J.~J.}\
  \bibnamefont {Slagmolen}}, \ and\ \bibinfo {author} {\bibfnamefont
  {K.}~\bibnamefont {Venkateswara}},\ }\href {\doibase
  10.1103/PhysRevLett.121.221104} {\bibfield  {journal} {\bibinfo  {journal}
  {Phys. Rev. Lett.}\ }\textbf {\bibinfo {volume} {121}},\ \bibinfo {pages}
  {221104} (\bibinfo {year} {2018}{\natexlab{b}})}\BibitemShut {NoStop}%
\bibitem [{\citenamefont {Tringali}\ \emph {et~al.}(2019)\citenamefont
  {Tringali}, \citenamefont {Bulik}, \citenamefont {Harms}, \citenamefont
  {Fiori}, \citenamefont {Paoletti}, \citenamefont {Singh}, \citenamefont
  {Idzkowski}, \citenamefont {Kutynia}, \citenamefont {Nikliborc},
  \citenamefont {Suchi{\'{n}}ski}, \citenamefont {Bertolini},\ and\
  \citenamefont {Koley}}]{Tringali_2019}%
  \BibitemOpen
  \bibfield  {author} {\bibinfo {author} {\bibfnamefont {M.~C.}\ \bibnamefont
  {Tringali}}, \bibinfo {author} {\bibfnamefont {T.}~\bibnamefont {Bulik}},
  \bibinfo {author} {\bibfnamefont {J.}~\bibnamefont {Harms}}, \bibinfo
  {author} {\bibfnamefont {I.}~\bibnamefont {Fiori}}, \bibinfo {author}
  {\bibfnamefont {F.}~\bibnamefont {Paoletti}}, \bibinfo {author}
  {\bibfnamefont {N.}~\bibnamefont {Singh}}, \bibinfo {author} {\bibfnamefont
  {B.}~\bibnamefont {Idzkowski}}, \bibinfo {author} {\bibfnamefont
  {A.}~\bibnamefont {Kutynia}}, \bibinfo {author} {\bibfnamefont
  {K.}~\bibnamefont {Nikliborc}}, \bibinfo {author} {\bibfnamefont
  {M.}~\bibnamefont {Suchi{\'{n}}ski}}, \bibinfo {author} {\bibfnamefont
  {A.}~\bibnamefont {Bertolini}}, \ and\ \bibinfo {author} {\bibfnamefont
  {S.}~\bibnamefont {Koley}},\ }\href {\doibase 10.1088/1361-6382/ab5c43}
  {\bibfield  {journal} {\bibinfo  {journal} {Classical and Quantum Gravity}\
  }\textbf {\bibinfo {volume} {37}},\ \bibinfo {pages} {025005} (\bibinfo
  {year} {2019})}\BibitemShut {NoStop}%
\bibitem [{\citenamefont {Badaracco}\ \emph {et~al.}(2020)\citenamefont
  {Badaracco}, \citenamefont {Harms}, \citenamefont {Bertolini}, \citenamefont
  {Bulik}, \citenamefont {Fiori}, \citenamefont {Idzkowski}, \citenamefont
  {Kutynia}, \citenamefont {Nikliborc}, \citenamefont {Paoletti}, \citenamefont
  {Paoli}, \citenamefont {Rei},\ and\ \citenamefont
  {Suchinski}}]{Badaracco_2020}%
  \BibitemOpen
  \bibfield  {author} {\bibinfo {author} {\bibfnamefont {F.}~\bibnamefont
  {Badaracco}}, \bibinfo {author} {\bibfnamefont {J.}~\bibnamefont {Harms}},
  \bibinfo {author} {\bibfnamefont {A.}~\bibnamefont {Bertolini}}, \bibinfo
  {author} {\bibfnamefont {T.}~\bibnamefont {Bulik}}, \bibinfo {author}
  {\bibfnamefont {I.}~\bibnamefont {Fiori}}, \bibinfo {author} {\bibfnamefont
  {B.}~\bibnamefont {Idzkowski}}, \bibinfo {author} {\bibfnamefont
  {A.}~\bibnamefont {Kutynia}}, \bibinfo {author} {\bibfnamefont
  {K.}~\bibnamefont {Nikliborc}}, \bibinfo {author} {\bibfnamefont
  {F.}~\bibnamefont {Paoletti}}, \bibinfo {author} {\bibfnamefont
  {A.}~\bibnamefont {Paoli}}, \bibinfo {author} {\bibfnamefont
  {L.}~\bibnamefont {Rei}}, \ and\ \bibinfo {author} {\bibfnamefont
  {M.}~\bibnamefont {Suchinski}},\ }\href {\doibase 10.1088/1361-6382/abab64}
  {\bibfield  {journal} {\bibinfo  {journal} {Classical and Quantum Gravity}\
  }\textbf {\bibinfo {volume} {37}},\ \bibinfo {pages} {195016} (\bibinfo
  {year} {2020})}\BibitemShut {NoStop}%
\bibitem [{\citenamefont {Badaracco}\ and\ \citenamefont
  {Harms}(2019)}]{Badaracco_2019}%
  \BibitemOpen
  \bibfield  {author} {\bibinfo {author} {\bibfnamefont {F.}~\bibnamefont
  {Badaracco}}\ and\ \bibinfo {author} {\bibfnamefont {J.}~\bibnamefont
  {Harms}},\ }\href {\doibase 10.1088/1361-6382/ab28c1} {\bibfield  {journal}
  {\bibinfo  {journal} {Classical and Quantum Gravity}\ }\textbf {\bibinfo
  {volume} {36}},\ \bibinfo {pages} {145006} (\bibinfo {year}
  {2019})}\BibitemShut {NoStop}%
\bibitem [{\citenamefont {Andric}\ and\ \citenamefont
  {Harms}(2020)}]{NN_Sardinia2020}%
  \BibitemOpen
  \bibfield  {author} {\bibinfo {author} {\bibfnamefont {T.}~\bibnamefont
  {Andric}}\ and\ \bibinfo {author} {\bibfnamefont {J.}~\bibnamefont {Harms}},\
  }\href {\doibase https://doi.org/10.1029/2020JB020401} {\bibfield  {journal}
  {\bibinfo  {journal} {Journal of Geophysical Research: Solid Earth}\ }\textbf
  {\bibinfo {volume} {125}},\ \bibinfo {pages} {e2020JB020401} (\bibinfo {year}
  {2020})},\ \bibinfo {note} {e2020JB020401 10.1029/2020JB020401}\BibitemShut
  {NoStop}%
\bibitem [{\citenamefont {Di~Giovanni}\ \emph {et~al.}(2020)\citenamefont
  {Di~Giovanni}, \citenamefont {Giunchi}, \citenamefont {Saccorotti},
  \citenamefont {Berbellini}, \citenamefont {Boschi}, \citenamefont {Olivieri},
  \citenamefont {De~Rosa}, \citenamefont {Naticchioni}, \citenamefont
  {Oggiano}, \citenamefont {Carpinelli}, \citenamefont {D’Urso},
  \citenamefont {Cuccuru}, \citenamefont {Sipala}, \citenamefont {Calloni},
  \citenamefont {Di~Fiore}, \citenamefont {Grado}, \citenamefont {Migoni},
  \citenamefont {Cardini}, \citenamefont {Paoletti}, \citenamefont {Fiori},
  \citenamefont {Harms}, \citenamefont {Majorana}, \citenamefont {Rapagnani},
  \citenamefont {Ricci},\ and\ \citenamefont {Punturo}}]{10.1785/0220200186}%
  \BibitemOpen
  \bibfield  {author} {\bibinfo {author} {\bibfnamefont {M.}~\bibnamefont
  {Di~Giovanni}}, \bibinfo {author} {\bibfnamefont {C.}~\bibnamefont
  {Giunchi}}, \bibinfo {author} {\bibfnamefont {G.}~\bibnamefont {Saccorotti}},
  \bibinfo {author} {\bibfnamefont {A.}~\bibnamefont {Berbellini}}, \bibinfo
  {author} {\bibfnamefont {L.}~\bibnamefont {Boschi}}, \bibinfo {author}
  {\bibfnamefont {M.}~\bibnamefont {Olivieri}}, \bibinfo {author}
  {\bibfnamefont {R.}~\bibnamefont {De~Rosa}}, \bibinfo {author} {\bibfnamefont
  {L.}~\bibnamefont {Naticchioni}}, \bibinfo {author} {\bibfnamefont
  {G.}~\bibnamefont {Oggiano}}, \bibinfo {author} {\bibfnamefont
  {M.}~\bibnamefont {Carpinelli}}, \bibinfo {author} {\bibfnamefont
  {D.}~\bibnamefont {D’Urso}}, \bibinfo {author} {\bibfnamefont
  {S.}~\bibnamefont {Cuccuru}}, \bibinfo {author} {\bibfnamefont
  {V.}~\bibnamefont {Sipala}}, \bibinfo {author} {\bibfnamefont
  {E.}~\bibnamefont {Calloni}}, \bibinfo {author} {\bibfnamefont
  {L.}~\bibnamefont {Di~Fiore}}, \bibinfo {author} {\bibfnamefont
  {A.}~\bibnamefont {Grado}}, \bibinfo {author} {\bibfnamefont
  {C.}~\bibnamefont {Migoni}}, \bibinfo {author} {\bibfnamefont
  {A.}~\bibnamefont {Cardini}}, \bibinfo {author} {\bibfnamefont
  {F.}~\bibnamefont {Paoletti}}, \bibinfo {author} {\bibfnamefont
  {I.}~\bibnamefont {Fiori}}, \bibinfo {author} {\bibfnamefont
  {J.}~\bibnamefont {Harms}}, \bibinfo {author} {\bibfnamefont
  {E.}~\bibnamefont {Majorana}}, \bibinfo {author} {\bibfnamefont
  {P.}~\bibnamefont {Rapagnani}}, \bibinfo {author} {\bibfnamefont
  {F.}~\bibnamefont {Ricci}}, \ and\ \bibinfo {author} {\bibfnamefont
  {M.}~\bibnamefont {Punturo}},\ }\href {\doibase 10.1785/0220200186}
  {\bibfield  {journal} {\bibinfo  {journal} {Seismological Research Letters}\
  }\textbf {\bibinfo {volume} {92}},\ \bibinfo {pages} {352} (\bibinfo {year}
  {2020})}\BibitemShut {NoStop}%
\bibitem [{\citenamefont {Bader}\ \emph {et~al.}(2022)\citenamefont {Bader},
  \citenamefont {Koley}, \citenamefont {van~den Brand}, \citenamefont
  {Campman}, \citenamefont {Bulten}, \citenamefont {Linde},\ and\ \citenamefont
  {Vink}}]{Bader_2022}%
  \BibitemOpen
  \bibfield  {author} {\bibinfo {author} {\bibfnamefont {M.}~\bibnamefont
  {Bader}}, \bibinfo {author} {\bibfnamefont {S.}~\bibnamefont {Koley}},
  \bibinfo {author} {\bibfnamefont {J.}~\bibnamefont {van~den Brand}}, \bibinfo
  {author} {\bibfnamefont {X.}~\bibnamefont {Campman}}, \bibinfo {author}
  {\bibfnamefont {H.~J.}\ \bibnamefont {Bulten}}, \bibinfo {author}
  {\bibfnamefont {F.}~\bibnamefont {Linde}}, \ and\ \bibinfo {author}
  {\bibfnamefont {B.}~\bibnamefont {Vink}},\ }\href {\doibase
  10.1088/1361-6382/ac1be4} {\bibfield  {journal} {\bibinfo  {journal}
  {Classical and Quantum Gravity}\ }\textbf {\bibinfo {volume} {39}},\ \bibinfo
  {pages} {025009} (\bibinfo {year} {2022})}\BibitemShut {NoStop}%
\bibitem [{\citenamefont {Koley}\ \emph {et~al.}(2022)\citenamefont {Koley},
  \citenamefont {Bader}, \citenamefont {van~den Brand}, \citenamefont
  {Campman}, \citenamefont {Bulten}, \citenamefont {Linde},\ and\ \citenamefont
  {Vink}}]{Koley_2022}%
  \BibitemOpen
  \bibfield  {author} {\bibinfo {author} {\bibfnamefont {S.}~\bibnamefont
  {Koley}}, \bibinfo {author} {\bibfnamefont {M.}~\bibnamefont {Bader}},
  \bibinfo {author} {\bibfnamefont {J.}~\bibnamefont {van~den Brand}}, \bibinfo
  {author} {\bibfnamefont {X.}~\bibnamefont {Campman}}, \bibinfo {author}
  {\bibfnamefont {H.~J.}\ \bibnamefont {Bulten}}, \bibinfo {author}
  {\bibfnamefont {F.}~\bibnamefont {Linde}}, \ and\ \bibinfo {author}
  {\bibfnamefont {B.}~\bibnamefont {Vink}},\ }\href {\doibase
  10.1088/1361-6382/ac2b08} {\bibfield  {journal} {\bibinfo  {journal}
  {Classical and Quantum Gravity}\ }\textbf {\bibinfo {volume} {39}},\ \bibinfo
  {pages} {025008} (\bibinfo {year} {2022})}\BibitemShut {NoStop}%
\bibitem [{\citenamefont {Coughlin}\ \emph {et~al.}(2019)\citenamefont
  {Coughlin}, \citenamefont {Harms}, \citenamefont {Bowden}, \citenamefont
  {Meyers}, \citenamefont {Tsai}, \citenamefont {Mandic}, \citenamefont
  {Pavlis},\ and\ \citenamefont {Prestegard}}]{Coughlin2019_seismic}%
  \BibitemOpen
  \bibfield  {author} {\bibinfo {author} {\bibfnamefont {M.}~\bibnamefont
  {Coughlin}}, \bibinfo {author} {\bibfnamefont {J.}~\bibnamefont {Harms}},
  \bibinfo {author} {\bibfnamefont {D.~C.}\ \bibnamefont {Bowden}}, \bibinfo
  {author} {\bibfnamefont {P.}~\bibnamefont {Meyers}}, \bibinfo {author}
  {\bibfnamefont {V.~C.}\ \bibnamefont {Tsai}}, \bibinfo {author}
  {\bibfnamefont {V.}~\bibnamefont {Mandic}}, \bibinfo {author} {\bibfnamefont
  {G.}~\bibnamefont {Pavlis}}, \ and\ \bibinfo {author} {\bibfnamefont
  {T.}~\bibnamefont {Prestegard}},\ }\href {\doibase 10.1029/2018jb016608}
  {\bibfield  {journal} {\bibinfo  {journal} {Journal of Geophysical Research:
  Solid Earth}\ }\textbf {\bibinfo {volume} {124}},\ \bibinfo {pages}
  {2941–2956} (\bibinfo {year} {2019})}\BibitemShut {NoStop}%
\bibitem [{\citenamefont {{ET Steering Committee Editorial
  Team}}(2020)}]{ETdesignRep}%
  \BibitemOpen
  \bibfield  {author} {\bibinfo {author} {\bibnamefont {{ET Steering Committee
  Editorial Team}}},\ }\href {https://apps.et-gw.eu/tds/?content=3&r=17245} {\
  (\bibinfo {year} {2020})},\ \Eprint {http://arxiv.org/abs/ET-0007B-20}
  {ET-0007B-20} \BibitemShut {NoStop}%
\bibitem [{\citenamefont {{Shahar Shani-Kadmiel}}\ \emph
  {et~al.}(2020)\citenamefont {{Shahar Shani-Kadmiel}}, \citenamefont {{Frank
  Linde}}, \citenamefont {{Läslo Evers}},\ and\ \citenamefont {{Bjorn
  Vink}}}]{3T_network}%
  \BibitemOpen
  \bibfield  {author} {\bibinfo {author} {\bibnamefont {{Shahar
  Shani-Kadmiel}}}, \bibinfo {author} {\bibnamefont {{Frank Linde}}}, \bibinfo
  {author} {\bibnamefont {{Läslo Evers}}}, \ and\ \bibinfo {author}
  {\bibnamefont {{Bjorn Vink}}},\ }\href {\doibase 10.7914/SN/3T_2020}
  {\enquote {\bibinfo {title} {Einstein telescope seismic campaigns},}\ }
  (\bibinfo {year} {2020})\BibitemShut {NoStop}%
\bibitem [{\citenamefont {Mandic}\ \emph {et~al.}(2018)\citenamefont {Mandic},
  \citenamefont {Tsai}, \citenamefont {Pavlis}, \citenamefont {Prestegard},
  \citenamefont {Bowden}, \citenamefont {Meyers},\ and\ \citenamefont
  {Caton}}]{10.1785/0220170228}%
  \BibitemOpen
  \bibfield  {author} {\bibinfo {author} {\bibfnamefont {V.}~\bibnamefont
  {Mandic}}, \bibinfo {author} {\bibfnamefont {V.~C.}\ \bibnamefont {Tsai}},
  \bibinfo {author} {\bibfnamefont {G.~L.}\ \bibnamefont {Pavlis}}, \bibinfo
  {author} {\bibfnamefont {T.}~\bibnamefont {Prestegard}}, \bibinfo {author}
  {\bibfnamefont {D.~C.}\ \bibnamefont {Bowden}}, \bibinfo {author}
  {\bibfnamefont {P.}~\bibnamefont {Meyers}}, \ and\ \bibinfo {author}
  {\bibfnamefont {R.}~\bibnamefont {Caton}},\ }\href {\doibase
  10.1785/0220170228} {\bibfield  {journal} {\bibinfo  {journal} {Seismological
  Research Letters}\ }\textbf {\bibinfo {volume} {89}},\ \bibinfo {pages}
  {2420} (\bibinfo {year} {2018})}\BibitemShut {NoStop}%
\bibitem [{\citenamefont {Koley}(2021)}]{SlidesSoumen}%
  \BibitemOpen
  \bibfield  {author} {\bibinfo {author} {\bibfnamefont {S.}~\bibnamefont
  {Koley}},\ }\href {\doibase 10.48550/ARXIV.2205.00416} {\enquote {\bibinfo
  {title} {Passive seismic with medium aperture arrays in limburg},}\ }
  (\bibinfo {year} {2021}),\ \Eprint
  {http://arxiv.org/abs/https://agenda.infn.it/event/28070/contributions/146753/}
  {https://agenda.infn.it/event/28070/contributions/146753/} \BibitemShut
  {NoStop}%
\bibitem [{\citenamefont {Peterson}(1993)}]{Pet1993}%
  \BibitemOpen
  \bibfield  {author} {\bibinfo {author} {\bibfnamefont {J.}~\bibnamefont
  {Peterson}},\ }\href@noop {} {\bibfield  {journal} {\bibinfo  {journal}
  {Open-file report}\ }\textbf {\bibinfo {volume} {93-322}} (\bibinfo {year}
  {1993})}\BibitemShut {NoStop}%
\bibitem [{\citenamefont {Yokoi}\ and\ \citenamefont
  {Margaryan}(2008)}]{Yokoi2008}%
  \BibitemOpen
  \bibfield  {author} {\bibinfo {author} {\bibfnamefont {T.}~\bibnamefont
  {Yokoi}}\ and\ \bibinfo {author} {\bibfnamefont {S.}~\bibnamefont
  {Margaryan}},\ }\href {\doibase 10.1111/j.1365-2478.2008.00709.x} {\bibfield
  {journal} {\bibinfo  {journal} {Geophysical Prospecting}\ }\textbf {\bibinfo
  {volume} {56}},\ \bibinfo {pages} {435} (\bibinfo {year} {2008})}\BibitemShut
  {NoStop}%
\bibitem [{\citenamefont {Saulson}(1984)}]{PhysRevD.30.732}%
  \BibitemOpen
  \bibfield  {author} {\bibinfo {author} {\bibfnamefont {P.~R.}\ \bibnamefont
  {Saulson}},\ }\href {\doibase 10.1103/PhysRevD.30.732} {\bibfield  {journal}
  {\bibinfo  {journal} {Phys. Rev. D}\ }\textbf {\bibinfo {volume} {30}},\
  \bibinfo {pages} {732} (\bibinfo {year} {1984})}\BibitemShut {NoStop}%
\bibitem [{\citenamefont {Harms}(2019)}]{2019}%
  \BibitemOpen
  \bibfield  {author} {\bibinfo {author} {\bibfnamefont {J.}~\bibnamefont
  {Harms}},\ }\href {\doibase 10.1007/s41114-019-0022-2} {\bibfield  {journal}
  {\bibinfo  {journal} {Living Reviews in Relativity}\ }\textbf {\bibinfo
  {volume} {22}} (\bibinfo {year} {2019}),\
  10.1007/s41114-019-0022-2}\BibitemShut {NoStop}%
\bibitem [{\citenamefont {Hughes}\ and\ \citenamefont
  {Thorne}(1998)}]{PhysRevD.58.122002}%
  \BibitemOpen
  \bibfield  {author} {\bibinfo {author} {\bibfnamefont {S.~A.}\ \bibnamefont
  {Hughes}}\ and\ \bibinfo {author} {\bibfnamefont {K.~S.}\ \bibnamefont
  {Thorne}},\ }\href {\doibase 10.1103/PhysRevD.58.122002} {\bibfield
  {journal} {\bibinfo  {journal} {Phys. Rev. D}\ }\textbf {\bibinfo {volume}
  {58}},\ \bibinfo {pages} {122002} (\bibinfo {year} {1998})}\BibitemShut
  {NoStop}%
\bibitem [{\citenamefont {Beker}\ \emph {et~al.}(2012)\citenamefont {Beker},
  \citenamefont {van~den Brand}, \citenamefont {Hennes},\ and\ \citenamefont
  {Rabeling}}]{Beker_2012}%
  \BibitemOpen
  \bibfield  {author} {\bibinfo {author} {\bibfnamefont {M.~G.}\ \bibnamefont
  {Beker}}, \bibinfo {author} {\bibfnamefont {J.~F.~J.}\ \bibnamefont {van~den
  Brand}}, \bibinfo {author} {\bibfnamefont {E.}~\bibnamefont {Hennes}}, \ and\
  \bibinfo {author} {\bibfnamefont {D.~S.}\ \bibnamefont {Rabeling}},\ }\href
  {\doibase 10.1088/1742-6596/363/1/012004} {\bibfield  {journal} {\bibinfo
  {journal} {Journal of Physics: Conference Series}\ }\textbf {\bibinfo
  {volume} {363}},\ \bibinfo {pages} {012004} (\bibinfo {year}
  {2012})}\BibitemShut {NoStop}%
\bibitem [{\citenamefont {Creighton}(2008)}]{creighton2008tumbleweeds}%
  \BibitemOpen
  \bibfield  {author} {\bibinfo {author} {\bibfnamefont {T.}~\bibnamefont
  {Creighton}},\ }\href@noop {} {\bibfield  {journal} {\bibinfo  {journal}
  {Classical and Quantum Gravity}\ }\textbf {\bibinfo {volume} {25}},\ \bibinfo
  {pages} {125011} (\bibinfo {year} {2008})}\BibitemShut {NoStop}%
\bibitem [{\citenamefont {Harms}\ and\ \citenamefont {Hild}(2014)}]{Harms2014}%
  \BibitemOpen
  \bibfield  {author} {\bibinfo {author} {\bibfnamefont {J.}~\bibnamefont
  {Harms}}\ and\ \bibinfo {author} {\bibfnamefont {S.}~\bibnamefont {Hild}},\
  }\href {\doibase 10.1088/0264-9381/31/18/185011} {\bibfield  {journal}
  {\bibinfo  {journal} {Classical and Quantum Gravity}\ }\textbf {\bibinfo
  {volume} {31}},\ \bibinfo {pages} {185011} (\bibinfo {year}
  {2014})}\BibitemShut {NoStop}%
\bibitem [{\citenamefont {Driggers}\ \emph
  {et~al.}(2012{\natexlab{b}})\citenamefont {Driggers}, \citenamefont {Evans},
  \citenamefont {Pepper},\ and\ \citenamefont {Adhikari}}]{Driggers2012}%
  \BibitemOpen
  \bibfield  {author} {\bibinfo {author} {\bibfnamefont {J.~C.}\ \bibnamefont
  {Driggers}}, \bibinfo {author} {\bibfnamefont {M.}~\bibnamefont {Evans}},
  \bibinfo {author} {\bibfnamefont {K.}~\bibnamefont {Pepper}}, \ and\ \bibinfo
  {author} {\bibfnamefont {R.}~\bibnamefont {Adhikari}},\ }\href {\doibase
  10.1063/1.3675891} {\bibfield  {journal} {\bibinfo  {journal} {Review of
  Scientific Instruments}\ }\textbf {\bibinfo {volume} {83}},\ \bibinfo {pages}
  {024501} (\bibinfo {year} {2012}{\natexlab{b}})}\BibitemShut {NoStop}%
\bibitem [{\citenamefont {Novotny}(1999)}]{novotny1999seismic}%
  \BibitemOpen
  \bibfield  {author} {\bibinfo {author} {\bibfnamefont {O.}~\bibnamefont
  {Novotny}},\ }\href@noop {} {\bibfield  {journal} {\bibinfo  {journal}
  {Bahia, Salvador: Instituto de Geociencias}\ }\textbf {\bibinfo {volume}
  {61}} (\bibinfo {year} {1999})}\BibitemShut {NoStop}%
\bibitem [{\citenamefont {Harms}\ \emph {et~al.}(2022)\citenamefont {Harms},
  \citenamefont {Naticchioni}, \citenamefont {Calloni}, \citenamefont
  {De~Rosa}, \citenamefont {Ricci},\ and\ \citenamefont
  {D'Urso}}]{harms2022lower}%
  \BibitemOpen
  \bibfield  {author} {\bibinfo {author} {\bibfnamefont {J.}~\bibnamefont
  {Harms}}, \bibinfo {author} {\bibfnamefont {L.}~\bibnamefont {Naticchioni}},
  \bibinfo {author} {\bibfnamefont {E.}~\bibnamefont {Calloni}}, \bibinfo
  {author} {\bibfnamefont {R.}~\bibnamefont {De~Rosa}}, \bibinfo {author}
  {\bibfnamefont {F.}~\bibnamefont {Ricci}}, \ and\ \bibinfo {author}
  {\bibfnamefont {D.}~\bibnamefont {D'Urso}},\ }\href@noop {} {\bibfield
  {journal} {\bibinfo  {journal} {arXiv preprint arXiv:2202.12841}\ } (\bibinfo
  {year} {2022})}\BibitemShut {NoStop}%
\bibitem [{\citenamefont {Bormann}\ \emph {et~al.}(2002)\citenamefont
  {Bormann}, \citenamefont {Engdahl},\ and\ \citenamefont
  {Kind}}]{BoEA2002ch2}%
  \BibitemOpen
  \bibfield  {author} {\bibinfo {author} {\bibfnamefont {P.}~\bibnamefont
  {Bormann}}, \bibinfo {author} {\bibfnamefont {B.}~\bibnamefont {Engdahl}}, \
  and\ \bibinfo {author} {\bibfnamefont {R.}~\bibnamefont {Kind}},\ }\enquote
  {\bibinfo {title} {{New Manual of Seismological Observatory Practice}},}\ \
  (\bibinfo  {publisher} {GFZ Potsdam},\ \bibinfo {year} {2002})\
  Chap.~\bibinfo {chapter} {2}\BibitemShut {NoStop}%
\bibitem [{\citenamefont {Hild}\ \emph {et~al.}(2009)\citenamefont {Hild},
  \citenamefont {Chelkowski}, \citenamefont {Freise}, \citenamefont {Franc},
  \citenamefont {Morgado}, \citenamefont {Flaminio},\ and\ \citenamefont
  {DeSalvo}}]{Hild_2009}%
  \BibitemOpen
  \bibfield  {author} {\bibinfo {author} {\bibfnamefont {S.}~\bibnamefont
  {Hild}}, \bibinfo {author} {\bibfnamefont {S.}~\bibnamefont {Chelkowski}},
  \bibinfo {author} {\bibfnamefont {A.}~\bibnamefont {Freise}}, \bibinfo
  {author} {\bibfnamefont {J.}~\bibnamefont {Franc}}, \bibinfo {author}
  {\bibfnamefont {N.}~\bibnamefont {Morgado}}, \bibinfo {author} {\bibfnamefont
  {R.}~\bibnamefont {Flaminio}}, \ and\ \bibinfo {author} {\bibfnamefont
  {R.}~\bibnamefont {DeSalvo}},\ }\href {\doibase
  10.1088/0264-9381/27/1/015003} {\bibfield  {journal} {\bibinfo  {journal}
  {Classical and Quantum Gravity}\ }\textbf {\bibinfo {volume} {27}},\ \bibinfo
  {pages} {015003} (\bibinfo {year} {2009})}\BibitemShut {NoStop}%
\bibitem [{\citenamefont {Christensen}(1992)}]{PhysRevD.46.5250}%
  \BibitemOpen
  \bibfield  {author} {\bibinfo {author} {\bibfnamefont {N.}~\bibnamefont
  {Christensen}},\ }\href {\doibase 10.1103/PhysRevD.46.5250} {\bibfield
  {journal} {\bibinfo  {journal} {Phys. Rev. D}\ }\textbf {\bibinfo {volume}
  {46}},\ \bibinfo {pages} {5250} (\bibinfo {year} {1992})}\BibitemShut
  {NoStop}%
\bibitem [{\citenamefont {Allen}\ and\ \citenamefont
  {Romano}(1999)}]{PhysRevD.59.102001}%
  \BibitemOpen
  \bibfield  {author} {\bibinfo {author} {\bibfnamefont {B.}~\bibnamefont
  {Allen}}\ and\ \bibinfo {author} {\bibfnamefont {J.~D.}\ \bibnamefont
  {Romano}},\ }\href {\doibase 10.1103/PhysRevD.59.102001} {\bibfield
  {journal} {\bibinfo  {journal} {Phys. Rev. D}\ }\textbf {\bibinfo {volume}
  {59}},\ \bibinfo {pages} {102001} (\bibinfo {year} {1999})}\BibitemShut
  {NoStop}%
\bibitem [{\citenamefont {Romano}\ and\ \citenamefont
  {Cornish}(2017{\natexlab{a}})}]{LivingRevRelativ20}%
  \BibitemOpen
  \bibfield  {author} {\bibinfo {author} {\bibfnamefont {J.~D.}\ \bibnamefont
  {Romano}}\ and\ \bibinfo {author} {\bibfnamefont {N.~J.}\ \bibnamefont
  {Cornish}},\ }\href {\doibase 10.1007/s41114-017-0004-1} {\bibfield
  {journal} {\bibinfo  {journal} {Living Rev. Relativ.}\ }\textbf {\bibinfo
  {volume} {20}},\ \bibinfo {pages} {2} (\bibinfo {year}
  {2017}{\natexlab{a}})}\BibitemShut {NoStop}%
\bibitem [{\citenamefont {Ade}\ \emph {et~al.}(2016)\citenamefont {Ade} \emph
  {et~al.}}]{Planck:2015fie}%
  \BibitemOpen
  \bibfield  {author} {\bibinfo {author} {\bibfnamefont {P.~A.~R.}\
  \bibnamefont {Ade}} \emph {et~al.} (\bibinfo {collaboration} {Planck}),\
  }\href {\doibase 10.1051/0004-6361/201525830} {\bibfield  {journal} {\bibinfo
   {journal} {Astron. Astrophys.}\ }\textbf {\bibinfo {volume} {594}},\
  \bibinfo {pages} {A13} (\bibinfo {year} {2016})},\ \Eprint
  {http://arxiv.org/abs/1502.01589} {arXiv:1502.01589 [astro-ph.CO]}
  \BibitemShut {NoStop}%
\bibitem [{\citenamefont {Romano}\ and\ \citenamefont
  {Cornish}(2017{\natexlab{b}})}]{Romano:2016dpx}%
  \BibitemOpen
  \bibfield  {author} {\bibinfo {author} {\bibfnamefont {J.~D.}\ \bibnamefont
  {Romano}}\ and\ \bibinfo {author} {\bibfnamefont {N.~J.}\ \bibnamefont
  {Cornish}},\ }\href {\doibase 10.1007/s41114-017-0004-1} {\bibfield
  {journal} {\bibinfo  {journal} {Living Rev. Rel.}\ }\textbf {\bibinfo
  {volume} {20}},\ \bibinfo {pages} {2} (\bibinfo {year}
  {2017}{\natexlab{b}})},\ \Eprint {http://arxiv.org/abs/1608.06889}
  {arXiv:1608.06889 [gr-qc]} \BibitemShut {NoStop}%
\bibitem [{\citenamefont {Thrane}\ and\ \citenamefont
  {Romano}(2013)}]{Thrane:2013oya}%
  \BibitemOpen
  \bibfield  {author} {\bibinfo {author} {\bibfnamefont {E.}~\bibnamefont
  {Thrane}}\ and\ \bibinfo {author} {\bibfnamefont {J.~D.}\ \bibnamefont
  {Romano}},\ }\href {\doibase 10.1103/PhysRevD.88.124032} {\bibfield
  {journal} {\bibinfo  {journal} {Phys. Rev. D}\ }\textbf {\bibinfo {volume}
  {88}},\ \bibinfo {pages} {124032} (\bibinfo {year} {2013})},\ \Eprint
  {http://arxiv.org/abs/1310.5300} {arXiv:1310.5300 [astro-ph.IM]} \BibitemShut
  {NoStop}%
\bibitem [{\citenamefont {{LIGO Scientific Collaboration, Virgo Collaboration
  and KAGRA Collaboration}}()}]{O3IsotropicDataset}%
  \BibitemOpen
  \bibfield  {author} {\bibinfo {author} {\bibnamefont {{LIGO Scientific
  Collaboration, Virgo Collaboration and KAGRA Collaboration}}},\ }\href
  {https://dcc.ligo.org/G2001287/public} {\enquote {\bibinfo {title} {Data for
  upper limits on the isotropic gravitational-wave background from advanced
  {LIGO}'s and advanced {Virgo}'s third observing run},}\ }\BibitemShut
  {NoStop}%
\bibitem [{\citenamefont {Abbott}\ \emph {et~al.}(2021)\citenamefont {Abbott}
  \emph {et~al.}}]{KAGRA:2021kbb}%
  \BibitemOpen
  \bibfield  {author} {\bibinfo {author} {\bibfnamefont {R.}~\bibnamefont
  {Abbott}} \emph {et~al.} (\bibinfo {collaboration} {KAGRA, Virgo, LIGO
  Scientific}),\ }\href {\doibase 10.1103/PhysRevD.104.022004} {\bibfield
  {journal} {\bibinfo  {journal} {Phys. Rev. D}\ }\textbf {\bibinfo {volume}
  {104}},\ \bibinfo {pages} {022004} (\bibinfo {year} {2021})},\ \Eprint
  {http://arxiv.org/abs/2101.12130} {arXiv:2101.12130 [gr-qc]} \BibitemShut
  {NoStop}%
\bibitem [{\citenamefont {Abbott}\ and\ \citenamefont
  {et~al}(2020)}]{APlussReference}%
  \BibitemOpen
  \bibfield  {author} {\bibinfo {author} {\bibfnamefont {B.~P.}\ \bibnamefont
  {Abbott}}\ and\ \bibinfo {author} {\bibnamefont {et~al}} (\bibinfo
  {collaboration} {KAGRA, Virgo, LIGO Scientific}),\ }\href {\doibase
  10.1007/s41114-020-00026-9} {\bibfield  {journal} {\bibinfo  {journal}
  {Living Reviews in Relativity}\ }\textbf {\bibinfo {volume} {23}} (\bibinfo
  {year} {2020}),\ 10.1007/s41114-020-00026-9}\BibitemShut {NoStop}%
\bibitem [{\citenamefont {Regimbau}\ \emph {et~al.}(2014)\citenamefont
  {Regimbau}, \citenamefont {Meacher},\ and\ \citenamefont
  {Coughlin}}]{PhysRevD.89.084046}%
  \BibitemOpen
  \bibfield  {author} {\bibinfo {author} {\bibfnamefont {T.}~\bibnamefont
  {Regimbau}}, \bibinfo {author} {\bibfnamefont {D.}~\bibnamefont {Meacher}}, \
  and\ \bibinfo {author} {\bibfnamefont {M.}~\bibnamefont {Coughlin}},\ }\href
  {\doibase 10.1103/PhysRevD.89.084046} {\bibfield  {journal} {\bibinfo
  {journal} {Phys. Rev. D}\ }\textbf {\bibinfo {volume} {89}},\ \bibinfo
  {pages} {084046} (\bibinfo {year} {2014})}\BibitemShut {NoStop}%
\bibitem [{\citenamefont {Maggiore}\ \emph {et~al.}(2020)\citenamefont
  {Maggiore} \emph {et~al.}}]{Maggiore:2019uih}%
  \BibitemOpen
  \bibfield  {author} {\bibinfo {author} {\bibfnamefont {M.}~\bibnamefont
  {Maggiore}} \emph {et~al.},\ }\href {\doibase 10.1088/1475-7516/2020/03/050}
  {\bibfield  {journal} {\bibinfo  {journal} {JCAP}\ }\textbf {\bibinfo
  {volume} {03}},\ \bibinfo {pages} {050} (\bibinfo {year} {2020})},\ \Eprint
  {http://arxiv.org/abs/1912.02622} {arXiv:1912.02622 [astro-ph.CO]}
  \BibitemShut {NoStop}%
\bibitem [{\citenamefont {{Janssens}}\ \emph {et~al.}(2022)\citenamefont
  {{Janssens}}, \citenamefont {{Boileau}}, \citenamefont {{Bizouard}},
  \citenamefont {{Christensen}}, \citenamefont {{Regimbau}},\ and\
  \citenamefont {{van Remortel}}}]{2022arXiv220500416J}%
  \BibitemOpen
  \bibfield  {author} {\bibinfo {author} {\bibfnamefont {K.}~\bibnamefont
  {{Janssens}}}, \bibinfo {author} {\bibfnamefont {G.}~\bibnamefont
  {{Boileau}}}, \bibinfo {author} {\bibfnamefont {M.-A.}\ \bibnamefont
  {{Bizouard}}}, \bibinfo {author} {\bibfnamefont {N.}~\bibnamefont
  {{Christensen}}}, \bibinfo {author} {\bibfnamefont {T.}~\bibnamefont
  {{Regimbau}}}, \ and\ \bibinfo {author} {\bibfnamefont {N.}~\bibnamefont
  {{van Remortel}}},\ }\href@noop {} {\bibfield  {journal} {\bibinfo  {journal}
  {arXiv e-prints}\ ,\ \bibinfo {eid} {arXiv:2205.00416}} (\bibinfo {year}
  {2022})},\ \Eprint {http://arxiv.org/abs/2205.00416} {arXiv:2205.00416
  [gr-qc]} \BibitemShut {NoStop}%
\end{thebibliography}%

\end{document}